# Identification of spin wave resonances and crystal field transitions in simple chromites RCrO$_3$ (R=Pr, Sm, Er) at ultralow temperatures in the THz spectral region

Néstor E. Massa*,[1] Karsten Holldack,[2] Rodolphe Sopracase,[3] Vinh Ta Phuoc,[3] Leire del Campo,[4] Patrick Echegut,[4] and José Antonio Alonso[5]

[1] Laboratorio Nacional de Investigación y Servicios en Espectroscopía Óptica-Centro CEQUINOR, Universidad Nacional de La Plata, 1900 La Plata, Argentina.

[2]Helmholtz-Zentrum für Materialien und Energie GmbH, Institut für Methoden und Instrumentierung der Forschung mit Synchrotronstrahlung (BESSYII) Berlin, Germany.

[3]CNRS- Groupe de Recherche en Matériaux, Microélectronique, Acoustique et Nanotechnologies, Université François Rabelais. 37200 Tours, France.

[4] CNRS, CEMHTI UPR3079, Université d'Orléans, F-45071 Orléans, France

[5]Instituto de Ciencia de Materiales de Madrid, CSIC, Cantoblanco, E-28049 Madrid, Spain.

*e-mail: neemmassa@gmail.com

# ABSTRACT

We report on THz absorption spectroscopy combined with high magnetic fields of polycrystalline $RCrO_3$ (R=Pr, Sm, Er) aiming understanding spin wave resonances at their low temperature magnetic phases.

Our measurements show that the temperature, and the implicit anisotropies at which the $Cr^{3+}$ spin reorientation at $T_{SR}$ takes place, are determinant on the ferromagnetic-like (FM) and the antiferromagnetic-like (AFM) spin modes being optically active. It is found that they are dependent on Rare Earth $4f$ moment and ion size.

We also studied temperature and field dependence of crystal field levels in the same spectroscopic region. $Pr^{3+}$ non-Kramers emerges at 100 K and Zeeman splits. An observed absence of spin wave resonances in $PrCrO_3$ is attributed to $Pr^{3+}$ remaining paramagnetic. In $SmCrO_3$ near cancelation of the spin and orbital moments is proposed as the possible reason for not detecting $Sm^{3+}$ ground state transitions. Here, the FM and AFM resonant modes harden when the temperature decreases and split linearly on applied fields at 5 K and below. In $ErCrO_3$ the $Er^{3+}$ Kramers doublet becomes active at about the $T_{SR}$ onset. Each line further experiences Zeeman splitting under magnetic fields while an spin reversal induced by a ~2.5 T field, back to the $\Gamma_4$ ($F_z$) from the $\Gamma_1$ phase at 2 K, produces a secondary splitting.

The 5 K AFM and FM excitations in $ErCrO_3$ have a concerted frequency-intensity temperature dependence and a shoulder pointing to the $Er^{3+}$ smaller ion size also disrupting the two magnetic sublattice approximation. Both resonances reduce to one when the temperature is lowered to 2 K in the $\Gamma_1$ representation. Our findings have important implications on the complex interplay in the magneto-electrodynamics associated with the Rare-Earth $4f$ – 3d transition metal spin coupling and the structural A site instabilities in perovskite multiferroics.

**Key words**

THz spectroscopy, spin-wave resonance, multiferroics, collective effects ferromagnetic, antiferromagnetic, and ferromagnetic resonances.

# INTRODUCTION

The possibility of developing materials with long-range magnetic order simultaneous to a polar state has resulted in an outburst of research around simpler oxides searching for optimization of the interplay between dielectric and magnetic properties. From them, Rare Earth orthoferrites and orthochromites are distorted perovskites with two magnetic ions to which much renovated attention has been paid due to the possible applications of their multiferroics properties.[1,2] Their crystalline structure is associated at room temperature to the orthorhombic $D_{2h}^{16}$–P$bnm$ space group (see diffraction patterns in the Supplementary Information, Fig. S1)[3] with four molecules per unit cell.[4] Recent measurements relying on neutron and high resolution X-ray diffraction point to a subtle departure into the P$na2_1$ space group.[5, 6, 7]

Generally, in orthorhombic chromites the transition metal sublattice orders antiferromagnetically at $T_{N(Cr)}$ ~100 K in the $\Gamma_4$ ($G_x$, $A_y$, $F_z$) irreducible representation. At ~ 30 K above the Néel temperature, there is a significant interplay between short- and long-range structural order where, still paramagnetic, the competing magnetic correlations in an incipient polar environment signal the onset of complex magnetic exchanges $M^{3+}$-$M^{3+}$; $M^{3+}$-$R^{3+}$; and $R^{3+}$-$R^{3+}$.[8,9] The t-e hybridization, due to local site distortion and cooperative octahedral rotation weighing ferromagnetic $t_3$-O-$e_0$ and antiferromagnetic $t_3$-O-$t_3$ contributions is thought responsible for triggering antiferromagnetism.[10] These compounds become weak ferromagnetic ($F_z$) from the canting spin of the transition metal along the $c$ axis.

 $Cr^{3+}$ antiferromagnetism may induce negative magnetization about the same temperatures consequence of the interaction of opposite $R^{3+}$ moment to $c$ $Cr^{3+}$ canted moment.[11]

The spontaneous magnetization changes on cooling from being $\underline{a}$ axis oriented to one for the antiparallel Rare Earth momentum along the $\underline{c}$ axis at a characteristic reorientation temperature $T_{SR}$ that depends on the coupling of the Rare Earth-Transition Metal moments. The new magnetic order emerges below the $T_N$ but well above the temperature at which the Rare Earth spin order. It depends fundamentally on antisymmetric coupling and Rare Earth-Transition metal exchange interaction.[8] The resulting order belongs to the $\Gamma_2$ ($F_x$, $C_y$, $G_z$) irreducible representation as a consequence of net 3d- and 4f- electron coupling in the $Cr^{3+}$ and $R^{3+}$ sublattices. It may be understood as an effective Cr-Rare Earth magnetic field driving the transition metal spin. This effective field is of the order, larger or smaller, of an actual applied external magnetic field, and thus, it may be overcome by the application of external field triggering spin reorientation.[12]

Spin reorientation in chromites has been found in $NdCrO_3$ (Ref.[13]); $SmCrO_3$ (Refs. [14, 15]); $GdCrO_3$ (Ref. [16]) ; $ErCrO_3$ (Ref. [17]) at $T_{SR}$ ~35 K ($T_N$ ~224 K); ~35 K ($T_N$ ~192 K); <14 K ($T_N$ ~170 K); < 25 K ($T_N$ ~133 K), respectively. It has also been proposed that these magnetic transitions are actually accompanied by structural changes lowering the lattice symmetry from orthorhombic to monoclinic.[18] Our temperature dependent far infrared reflectivity suggesting such lattice changes as new phonons emerging on cooling is shown in Fig.S2.[3]

We are motivated by the complex magnetic order and electron correlations originating in the lanthanide's 4f occupation associated with the transition metal $Cr^{3+}$ that, at very low temperatures, have remained largely unexplored. We discuss free-ion multiplets of rare earths split by crystal field effects resulting in singlets for ions with the even number ($Pr^{3+}$, singlet $4f^2$) or in Kramers doublets for odd number of 4f electrons as in $Sm^{3+}$ (doublet $4f^5$) and $Er^{3+}$ (doublet $4f^{11}$).

We found that, while the compounds cool down below $T_{N(Cr)}$ with the Cr-Rare Earth interaction increasingly dominant, two main features appear in the absorption spectra. One of them is traced to optical $R^{3+}$ ground state transitions, which have a distinctive strength with peak positions changing little when the sample cools down, and a second set, at lower frequencies, composed by much weaker bands with stronger temperature dependence that are identified as Brillouin zone center spin wave resonant modes. They arise in magnetically ordered materials excited by the magnetic field $\underline{H}$ of the electromagnetic THz radiation coupling spin oscillations. We also argue below, as for our measurements on $ErCrO_3$, on Rare Earth induced local magnetic and structural distortion at non-equivalent A cage perovskite sites.

Understanding spin resonances in antiferromagnetic materials was first tackled by Keffer and Kittel.[19] In their approach, the four non-equivalent magnetic $Cr^{3+}$ sites of the orthorhombic structure are reduced to two magnetic sublattices. Each sublattice magnetization, $M_1$ and $M_2$, has its own equation of motion linked to the exchange interaction describing unequal precessions. The resonant energies are given by

$$\omega_{Res} = \gamma H_0 \pm \gamma [H_A \cdot (2 \cdot H_E + H_A)]^{1/2} \qquad (1)$$

where $\underline{H_A}$ is the anisotropy field, $\underline{H_E}$ ($H_{Ei} = \lambda M_i$, $i=1,2$) are exchange fields and $\gamma = ge/2mc$ the magneto mechanical factor with the gyromagnetic ratio $g$ as the spectroscopic splitting factor. One main conclusion is that even in absence of an external field $\underline{H_0}$ there are resonances due to the exchange and anisotropy fields. $\underline{H_0}$ breaks the symmetry branching into upper and lower energy modes meaning that if $M_1$ is close parallel to the applied field the effective field of $M_2$ has the opposite sign. This causes the known resonant energies diverging picture splitting from each other under an increasing external field $\underline{H_0}$.[19]

To gain more insight on the origin of these resonances it is necessary to adopt a more quantitative view of the magnetic Hamiltonian accounting three fundamental contributions, namely, exchange, anisotropy, and Zeeman terms. These are represented by the Heisenberg term. $\cdot J \cdot \sum_{i,j} S_i \cdot S_j$ taking care of the Coulomb interaction and the spin Pauli exclusion. The exchange integral, J, is positive for ferromagnets and negative for antiferromagnets. Secondly, the canting of a metal transition moment out of the $\underline{ab}$ plane true antiferromagnetism. $Cr^{3+}$ is also involved in another important exchange interaction known as Dzyaloshinkii-Moriya (DM) $D \cdot S_i \times S_j$ linking the crossed product of coupled spins. Earlier pointed out by Moriya[20] arises from the perturbative role of spin-orbit coupling on the $Cr^{3+}$ ground state. The resulting net effect is described by the non-Heisenberg term as a Dzyaloshinkii[21] field laying along the $\underline{c}$ axis weakly ferromagnetic moment. In addition, it is also necessary to consider contributions such as, $K_{eff} \cdot S_i$ due to spin configurations along a particular crystallographic direction promoting magnetic anisotropies associated to a constant defining the prevailing spin relative to the spatial direction.

On the other hand, since the consequences of the weaker structural departure from centrosymmetric *Pbnm* to non-centrosymmetric *Pna*$2_1$ space group has not yet been theoretically explored, away from a pure antisymmetric exchange interaction, we will keep the subject outside our data analysis. It reflects on how to weigh against symmetric the Moriya spin-orbit coupling and the associated Dzyaloshinkii field in $RCrO_3$. This is an early issue addressed by Treves[22] to determine, either as due to single ion magnetic anisotropy or to antisymmetric exchange interaction, the origin of spin canting in orthorhombic ferrites. It was concluded the antisymmetric exchange mechanism was predominant.[22]

Within that framework, that amounts to a long-range magnetic approach, we will also keep out very recent studies on effect of oxygen spin polarization on magnetic states. This is a view beyond the Heisenberg model in which the spin polarization of oxygen is taken into account. It contrasts the standard approach that by handling only transition metal interactions the oxygen effects are considered a property of the transition metal.[23]

Thus, the spin Hamiltonian for our compounds may next be written taking into account a single isotropic exchange constant coupling nearest neighbor transition metal spins, a single antisymmetric (canting) exchange constant, and two anisotropy constants.[24, 25, 26] .This reduces to

$$H_{spin} = 2 \cdot J \cdot \sum_{i,j} S_i \cdot S_j + \sum_{i,j}(D \cdot S_i \times S_j) + \sum_{i,j}(K_{eff} \cdot S_i)^2. \quad (2)$$

The ensuing spin wave eigenvalues at $\underline{k}$=0 are

$$\omega_{FM} = \{24 \cdot J \cdot S[2(K_a - K_c) \cdot S]\}^{½} \quad (3)$$

$$\omega_{AFM} = \{24 \cdot J \cdot S [6 \cdot D \cdot S \cdot \tan \beta + 2 \cdot K_a \cdot S]\}^{1/2} \quad (4)$$

where one more time, $J$ and $D$ are the isotropic Heisenberg and D the antisymmetric DM) exchange constants, $S$ is the spin moment of the $i^{th}$ and $j^{th}$ nearest-neighbor of the metal transition ions, and $K_a$ and $K_c$, are anisotropies along the $\underline{a}$ and $\underline{c}$ axes. Here β is the Cr momentum canting angle off the _**ab**_ plane. The exchanges taking into account the transition metal and rare-earth interactions.[26, 26]

The two spin wave resonances, AFM and FM, are named antiferromagnetic mode and the ferromagnetic mode because they depend on anisotropies in the _**ac**_ plane along the antiferromagnetic axis while the other aligns along the ferromagnetic _**c**_ axis, respectively.[25]

The coupling of the electromagnetic wave *H* field with the spin wave takes place at resonances near the Brioullin zone center. Thus, we expect from our polycrystalline samples a comparable pair of temperature and magnetic field dependent excitations with relative weaker but near equal intensity. Suggested by eqs. (3) and (4), and shown below, with the AFM mode at slighted higher energies than the corresponding FM band.

Earlier emphasis on optical studies was on ambient temperature isomorphous orthoferrites $RFeO_3$ (R=Rare Earth) using methods that still remain an adequate approximation for unraveling low-energy spin-wave absorptions.[25,27] Among others, White et al[24] reported their temperature dependent Raman scattering in $YFeO_3$, $SmFeO_3$, $DyFeO_3$, $HoFeO_3$, and $ErFeO_3$ and Koslov et al pioneered THz techniques measuring $RFeO_3$ (R=Y, Tm, Dy, Gd, Ho, Er, Tb).[28] More recently, Constable et al[27] studied temperature-dependent spin waves and neutron scattering in antiferromagnetic $NdFeO_3$, Fu et al[29] measured spin resonances in $SmFeO_3$, and Zhang et al characterized them in $TmFeO_3$ using terahertz time domain spectroscopy.[30]

We present below what to our knowledge is the first report on Rare Earth crystal field multiplets and spin wave resonances in polycrystalline orthochromites $RCrO_3$ (R=Pr, Sm, Er) measured at ultralow temperatures in the THz spectral region. These compounds may be considered good representatives of the complete $RCrO_3$ (R=Rare Earth) family. Since in optical measurements of polycrystalline samples all the information is obtained simultaneously, and excitations associated with magnetic dipoles are somehow ambiguous relative to those of an electric dipole, we also measured their respective absorbance under magnetic fields up to 10 T. These, made possible by the beam quality of low α-mode running at the HZB-BESSY II THz beamline, aim to add understanding on the origin and detection of the ferromagnetic and antiferromagnetic resonant modes as well as on Rare Earth splitting of ground multiplets.

## EXPERIMENTAL DETAILS

$RCrO_3$ (R= Pr, Sm, Er) polycrystalline samples were prepared by standard ceramic synthesis procedures. Stoichiometric amounts of analytical grade $Pr_6O_{11}$, $Sm_2O_3$ or $Er_2O_3$ and $Cr_2O_3$ powder oxides were thoroughly ground and heated in air at 1000ºC for 12 h and 1300ºC for 12 h in alumina crucibles. The purity of the samples was checked by X-ray diffraction. Then, pellets of 12 mm diameter were prepared by uniaxial pressing the raw powders and sintering the disks at 1300ºC for 2 h.

Neutron powder-diffraction diagrams for $PrCrO_3$ and $ErCrO_3$ were collected at the Paul Scherrer Institute (Switzerland). X-ray powder diffraction data for $SmCrO_3$ structure refinement were collected at room temperature with Cu K radiation. Shown in Fig. S1,[3] all data were analyzed using the Rietveld method and the refinements were carried out with the program FULLPROF. Magnetic THz runs were done in a Bruker IFS125 HR spectrometer. We collected a 120 average number of scans with up to 0.2 $cm^{-1}$ resolution and with a 6 μm multilayer Mylar-beamsplitter providing a window (together with z-cut Quartz windows in the magnet) between 10 $cm^{-1}$ and 120 $cm^{-1}$ in combination with ultra-sensitive liquid helium cooled Si bolometers (4.2 K and 1.6 K from Infrared Labs). This facility is located at the THz beamline at the electron storage ring BESSY

II in the Helmholtz-Zentrum Berlin (HZB). Regular and low-alpha modes of the storage ring were used. In the synchrotron low-alpha mode electrons are compressed within shorter bunches of only ~2 ps duration allowing far-infrared wave trains up to mW average power to overlap coherently in the THz range below 50 cm$^{-1}$ (Ref. **31**). Magnetic field measurements have been done using a superconducting magnet (Oxford Spectromag 4000, -10 T to +10 T) interfaced with the interferometer and at a beam path fully in high vacuum of ~10$^{-5}$ mbar. Some magnetic measurements were taken in the zero field cooled mode using an Oxford Optistat cryostat in the sample compartment of the IFS125 HR.

The reported absorbance spectra were normalized using as reference the same pellet transmission at temperatures at which there was only a flat response. Except explicitly mentioned measurements were performed at 2 K-20 K range. The temperature was measured with a calibrated Cernox Sensor from LakeShore Cryotronics mounted to the cooper block that holds the sample in the Variable Temperature Insert (VTI) of the Spectromag 4000 Magnet.

Near normal incidence temperature dependent infrared reflectivity spectra, Fig. S2,[3] were measured with 2 cm$^{-1}$ resolution in a Bruker IFS 66 v/S interferometer interfaced with a closed cycle He refrigerator. A layer of gold was applied on the sample in situ and reassessed based on the temperature. The data were then used as a reference to calculate mirror reflectivity to take into account the scattering of light on the sample surface. These last measurements were done at the Groupement de Recherche Matériaux Microélectronique Acoustique Nanotechnologies-Université François Rabelais Tours, Faculté des Sciences & Techniques, Tours, France.

## Results and Discussion

### i) PrCrO$_3$

PrCrO$_3$ is a Rare Earth orthochromite that is currently reported crystallizing at room temperature in the P*bnm* space group[4,32] distorted perovskite structure with four non-equivalent Cr$^{3+}$ and R$^{3+}$ ions per unit cell (Fig. S1). Cr$^{3+}$ orders antiferromagnetic at ~239 K. It also develops below T$_N$ weak ferromagnetism along the *c* axis due to a canting angle α =18± 1 (mrad) relative to the *a* oriented Cr$^{3+}$ sublattice moments.[33] This, however, does not materialize in a complete spin flit, by Cr$^{3+}$ triggering a spin reorientation transition, because Pr$^{3+}$ (singlet 4f$^2$) moments remain paramagnetic down to 1.5 K preventing meaningful constructive exchanges.[34,35] It is known that in lighter lanthanides, crystal fields are responsible for single ion anisotropy that result relevant to the lack of magnetic order.[36]

We found an absorption band gradually emerging at 33 cm$^{-1}$ and ~100 K in the antiferromagnetic phase (Fig. 1a). It is about the energy associated with the first excited state of the $^3$H$_4$ manifold created by spin-orbit and Coulomb interactions in Pr$^{3+}$ two 4*f* electrons.[37]

At 5 K, where there is a tendency to magnetic order in R$^{3+}$ (Ref. **38**), our ~33 cm$^{-1}$ band is well represented by two Lorentzians. They entail to a main strong feature on a top of a much weaker side band also observed by inelastic neutron scattering. The temperature dependent strongest band represents the thermal evolution at energies of the transition from a ground state singlet to the next higher excited singlet that associated with the very weak shoulder constitute an accidental doublet.[39]

The temperature dependence of the 33 cm$^{-1}$ band mimics emerging new phonons bands in the far infrared reflectivity spectra at frequencies where octahedral vibrational modes are altered by small displacement of the Cr$^{3+}$ ions (Fig. S2).[3] The new phonon detection starts at temperatures above $T_{N(Cr)}$ meaning an environment locally lacking the inversion center but supporting structural deviations in a non-centrosymmetric ferroelectric polar distorted phase.[2]

It is also worth emphasizing that down to 3 K, and at difference with what it is found for $SmCrO_3$ and $ErCrO_3$, we did not observe any feature that might be associated with the antiferromagnetic or the ferromagnetic spin resonant modes.

Cooling $PrCrO_3$ below $T_{N(Cr)}$ triggers a weak antiferromagnetic component along the *c* axis out of a perfect in-plane magnetic alignment. This is accounted by an effective field (Dzyaloshinkii field) along the ferroelectric axis. It also implies a magnetic bias for the Rare Earth moment that, if the feedback of the Rare Earth magnetization ($Pr^{3+}$) on the $Cr^{3+}$ canting is negligible, may be regarded constant.[40] On the other hand, constructive magnetic coupling due the simultaneous ordering of Cr and Pr moments require both belonging to the same representation. In our case, it is $\Gamma_4$ ($F_z$). This condition is not met if the Rare Earth remains paramagnetic and/or if there is a subtle lower symmetry by lattice distortion out of centrosymmetry and our hypothesis of identical crystallographic and magnetic unit cells.

Optimizing the DM interaction by increasing canting on cooling will be doubtful because $Pr^{3+}$ remaining paramagnetic at the lowest measured temperature disturbs the *c* magnetic coupling, and thus, the overall optically active spin wave modes. A way to visualize the consequences of the dislocating magnetic interaction is to recall that the spin resonant ferromagnetic mode, $\omega_{FM}$, receives its name due to the now distorted low frequency cone defined by the ferromagnetic moment precessing the *z* axis. For the antiferromagnetic resonant mode, $\omega_{AFM}$, the lack of the $Pr^{3+}$ magnetic ordering will also be disturbing since it is built on x,y out-of-phase oscillations and in-phase oscillations along the *z* axis (this is illustrated in the Supplementary Information, Fig. S3).[3, 26] We conclude that the interactions between the anisotropy field acting on the $Cr^{3+}$ ions under the $Pr^{3+}$ paramagnetic random fluctuations prevents a complete spin reorientation.[40] The null result for $\omega_{FM}$ and $\omega_{AFM}$ resonance detection would then be likely due to dynamically blurred oscillation averaging.

To add to the characterization of our crystal field excitation at 33 cm$^{-1}$ we also recorded its spectra under magnetic fields in 1T increments up 10 T (Fig. 1b). As shown in Fig. 2 upon applying 1 T, the main band narrows and increases in relative intensity. At 2 T and further, the peak positions undergo a Zeeman split (Fig. 2, inset) into a doublet passing from linear to quadratic as band profiles undergo strong broadening and reduction in relative intensities (Fig. 1b, inset).

The departure from linear energy split is generally understood proposing a Hamiltonian considering free and crystal field coupling and much weaker contributions from hyperfine, nuclear quadrupole, electronic, and nuclear Zeeman interaction.[41] The quadratic field dependency is associated to the effective Hamiltonian of the last four contributions being significant only for non-Kramers ions, where there is only a singlet electronic ground state. It emerges due to the electronic Zeeman interaction consequence electronic crystal field wave functions mixing by the applied magnetic field.[42] This is also at the origin of Van Vleck paramagnetism that is a small contribution for partially filled 4*f* ions only to be considered in ions with non-magnetic ground state such as our non-Kramers $Pr^{3+}$. Its consequence is a very weak net up shift at higher fields of crystal field level traces.

Peak positions from $Pr^{3+}$ fits (Fig. 2) show that in addition to the quadratic field dependence there is an unaccounted contribution upshifting the overall picture at the highest fields. We speculate that this is consequence of applying stronger magnetic fields distorting orbitals and thus changing local lattice symmetries. As it was stated above, chromites are better described at the local level by the Pna2$_1$ space group meaning losing the center of inversion by Rare Earth displacement relative to the *c* axis. The small distortion locally belongs to the non-centrosymmetric point group C$_s$ instead of orthorhombic centrosymmetric D$_{2h}$. In turn, this triggers a convoluted Zeeman split due to Rare Earth at the different sites each individually contributing to the broadening of spectra with non-equivalent split levels following increments of the applied magnetic field. Our data suggest a mixed scenario product of unaccounted events destroying the expected mirror image in trace levels. Peak positions at weaker fields are results of clean Lorenztian fits turning upon applying stronger fields, up to 10 T, into non-linear broadening for features with much lower oscillator strengths in a broader energy span (Fig. 2) suggesting an still not yet investigated magneto-dynamical field response at the local lattice microscopic level.[43] It is also worth noting that the spectra shown in the figure optimize band peak position. If instead it is chosen to stress the extra broadening, associated with the reduction of site symmetry at high fields, we found the same quality fits by using fairly distorted Lorenztians. Examples of these are shown in Fig (S4) and are used to define the confidence band in Fig. 2 (center inset).

## ii)    SmCrO$_3$

At T$_N$ ~ 197 K  SmCrO$_3$ is a non-centrosymmetric orthorhombic distorted perovskite (P*na*2$_1$) after structural instabilities tied to octahedral distortions and rare earth displacements leading to the onset of ferroelectricity.[44] Raman and infrared data confirm lattice parameter changes by local Cr off-center octahedra  at ~30 K above T$_N$ modifying vibrational  band profiles. It implies striction in the Cr$^{3+}$-Sm$^{3+}$ exchange and symmetry breaking contributing to polar polarization that also induces significant changes at the spin reorientation temperature $T_{SR}$.[44] At ~40 K, and below, *a* parallel $Cr^{3+}$ moments rotate realigning along the *c* axis antiparallel to $Sm^{3+}$ momenta in a second order transition. This change takes place within a 20 degrees interval under strong Sm-Cr hybridization[45] in which spins trigger a distinctive specific heat anomaly while cooling from $\Gamma_4$ ($G_x$, $A_y$, $F_z$) to $\Gamma_2$ ($F_x$, $C_y$, $G_z$).[46,47] This leads to the antiferromagnetically ordered interplay between $Sm^{3+}$ and $Cr^{3+}$ (Ref. **48**). In addition, at $T_{SR}$ spin-phonon interactions have a role in the ferroelectric polar distortion. Cr-Sm lattice distances coexisting with magnetoelastic and electroelastic couplings change discontinuously.[44,49]

Against this background, we do not observe any feature that might be associated with crystal field excitations (CEF) of $Sm^{3+}$. It is expected that crystal-field energy levels split in three Kramers doublets the $4f^5$ electronic configuration of $Sm^{3+}$ in the Cs crystal-field symmetry of the $Sm^{3+}$ sites.[50] We also found the same in our recent measurements of $SmNiO_3$ and $SmFeO_3$ concluding that it is reflection of an intrinsic property of Sm at the orthorhombic lattice site. A lack of crystal field excitations might relate to the near canceling of the spin and orbital moments found for isomorphous $SmCoO_3$ in the $^6H_{5/2}$ multiplet (L=5, S=5/2, J=L – S=5/2, $g_L$=2/7), which is manifested by low $g_{j'}$ factor of the pseudospins in the ground Kramers doublet.[51] On the other hand, earlier, $^6H_{5/2}$ ground-state excited levels at ~127 $cm^{-1}$ and ~234 $cm^{-1}$ were suggested from infrared measurements of $SmMnO_3$ from side bands at 100 K. Based on that and his inelastic neutron scattering, Ma then assigned a wide shoulder around ~161 $cm^{-1}$ (~20meV) and a broad peak around ~242 $cm^{-1}$ (~ 30meV) to crystal field effects (CFE) of $Sm^{3+}$ in isomorphous $SmFeO_3$ in spite of being Sm one of the strongest neutron absorber.[52] In this case, the CEF of Sm in $Sm^{3+}CrO_3$ would lay outside our measuring range and at odds with our findings for Pr and Er.

Strong moment clamping between local moment of $Sm^{3+}$ and spin reoriented $Cr^{3+}$ locked through oxygen ligands may add a bonus allowing a relative increase in coherence in the $Cr^{3+}$ spin precessions.

Fig. 3 shows the temperature dependence of spin wave mode absorbance in the low temperature $\Gamma_2$ ($F_x$, $C_y$, $G_z$) irreducible group of $SmCrO_3$. We find that the two resonances, allowed by effective field exchange and anisotropy fields, are well defined only below $T_{SR}$. These are the lower frequency FM mode due to out of phase z oscillations and AFM depending on *ac* plane anisotropies.[25,26] At increasing temperatures, the signal disappears within the background as if the relative stronger fluctuations introduced by order-disorder overcome the resonant modes. These spin wave modes close reproduce in $SmCrO_3$ the behavior found at the lowest temperature reported by Fu et al for $SmFeO_3$ (Ref. **29**). As temperature decreases band profiles become better defined and both modes harden. Since the temperatures in which the bands are best defined are just below the interval in which the spin reorientation is reported one may interpret the actual detection of resonant modes only happening when moment coherence is achieved under the same representation, which is, after some degree of $Cr^{3+}$ and $Sm^{3+}$ aligning. As in earlier conclusions, it points to the role of the Rare Earth-Metal transition superexchange[4, 53] at bands getting sharper and narrower as the sample cools down (Fig. 3). The difference in onset temperature between $SmCrO_3$ and $SmFeO_3$ is at unison with the magnitude of the Rare-Earth-transition-metal coupling being at least twice as great in orthochromites than in orthoferrites where the total spin is $S=5/2$ ($Fe^{3+}$: $t3e2$) (while for $SmCrO_3$ it is $S=3/2$ ($Cr^{3+}$: $t3\ e0$)). Transition metal ordering temperatures are a factor of two to six lower in orthochromites.[12]

For sake of completeness we also fit our data with a power law proposed in for SmFeO$_3$.[29] In our case, this yields

$$\omega_{AFM} = 0.6225 + (1.6868/T - (1.9221/T^{**}2) - (0.0145^*T) \quad \text{eq (5)}$$

$$\omega_{FM} = 0.6894 + (1.2384/T) - (1.3556/T^{**}2) - (0.0295^*T) \quad \text{eq (6)}$$

where $\omega_{AFM}$ and $\omega_{FM}$ are the resonant frequencies peak position in THz; $y_0$, a, b, c, are fitting constants, and T is the temperature. In the range in which the resonances were detected these results show about the same behavior as found for ferrite SmFeO$_3$, i.e., quasi-linear for the lowest temperature points departing to a weak power law at higher temperatures. Accordingly, it also suggests temperature dependent effective anisotropy constants with overall similar behaviors. [29]

The magnetic origin of the resonant profiles at 5 K is confirmed by their dependence on applied magnetic fields as it is shown in Fig. 4(a,b). The increment of the applied magnetic field is followed by further linear branch splitting up to 10 T as predicted by Keffer and Kittel.[19] It should be noted that band peak determination by Lorentzian fits at those higher fields is rather arbitrary due to bands becoming too broad together with an striking attenuation due to changes in optical and spin lifetimes. Intrinsic fluctuations, not contemplated in calculations, involve electrons in a lower than centrosymmetric environment in field induced distorted d-orbitals magneto-electrically coupled in the already intrinsic unstable perovskite lattice.[45]

*iii)*     *ErCrO$_3$*

ErCrO$_3$ undergoes antiferromagnetic ordering at T$_{N(Cr)}$~ 133K with spin reorientation starting at T$_{SR}$~ 22 K cooling from the $\Gamma_4$ (G$_x$, A$_y$, F$_z$) to the $\Gamma_1$(A$_x$,G$_y$,C$_z$) representation. The easy axis in ErCrO$_3$ rotates by 90º within the temperature range between 25 K and 10 K in a scenario of R$^{3+}$ spins increasing order and with the exchange Cr$^{3+}$- R$^{3+}$ optimizing the M$^{3+}$ realignment.[54] It was found that the intermediate spin configuration $\Gamma_2$ (F$_x$, C$_y$, G$_z$) may be retrieved starting from $\Gamma_1$(A$_x$,G$_y$,C$_z$) at low temperatures and applying a magnetic field that, if strong enough, may completely spin reorient.[55] Applied fields at about 2 T also lead to reversal of the electric polarization by coupling magnetization and polarization.[47]

In our spectral region a distinctive set of bands centered at 46 cm$^{-1}$ emerges on cooling between 100 K and 25 K clearly delineating the odd electron Kramers doublet of the Er$^{3+}$(4f$^{11}$) valence configuration. This, Fig. 5(a), is identified as the two lowest lying doublet of the $^4$I$_{15/2}$ ground state of ErCrO$_3$ (Ref 17). Neutron diffraction measurements show that Er$^{3+}$ ions order at 16.8 K simultaneously to Cr$^{3+}$ spin reorienting from magnetic G$_x$ to a combination of G$_y$ and G$_x$ modes below that temperature.[56, 57]

The Kramers doublet becomes optically active well above 25 K, suggesting that at these relative higher temperatures in ErCrO$_3$ the anisotropies discussed for SmCrO$_3$ are already present as Er$^{3+}$-Cr$^{3+}$ and Er$^{3+}$-Er$^{3+}$ exchanges in the polar lattice.[58] At 3 K, in the $\Gamma_1$ phase, the Kramers-doublet is the only lowest lying pair significantly occupied.[17] Under low applied fields each component of the doublet splits linearly up to a critical value H$_c$. Band profiles again can satisfactorily be deconvoluted by a suitable number of Lorentzians (Fig. 6). At H$_c$ ~2.5 T and higher these bands gradually distort into resonant asymmetric profiles that are also significantly broadened by non-equivalent rare earth ion sites contributing individually to superposed split energy levels. At the higher frequency end they have a sharp cut off that can only be best fitted by Weibull distributions (see, Fig. S5). Shown in Fig. 6 (inset), the new absorption line splits linearly at $H_c$ as the applied field overcomes the Cr-Rare Earth magnetic interaction. Its area grows gradually as we go from the $\Gamma_1$ phase to $\Gamma_4$. We also detect at ~50 cm$^{-1}$ an anomalous band broadening from which is possible to deconvolute (dashed lines under the bell shape band in fig. S5) and assign to the split of the lower Kramers energy level. The overall behavior is reminiscent to slope breaks reported by Toyokawa et al[45] at around 19000 cm$^{-1}$ as the field applied below $T_{SR}$ reverts $Cr^{3+}$ spins into the $\Gamma_4$ ($G_x$, $A_y$, $F_z$) phase in a first order phase transition. Disparity with earlier $H_c$ values (also Kaneko et al[60]) and claim of sharp onset at ~9 K are attributed to differences in the working techniques, and particularly, in the spectral resolution. In our measurements at ~9 K the Kramers doublet is better defined but the precursor onset, that in the strip-chart recording from grating spectrometers earlier days went undetected, takes place at a higher temperature.[17, 59, 60, 61] The new populated levels generated by the spin back rotation coexist with a polar distortion (ferroelectricity).[48]

On cooling in a parallel development with the activity of crystal field transitions, suggesting common driving mechanism within an order-disorder phase transition at ~25 K, two weaker features (Fig 7(a)) emerge whose intensities and frequencies change with temperature in a concerted way. We assign the band at ~26 cm$^{-1}$ to the ferromagnetic and that at ~ 42 cm$^{-1}$ to the antiferromagnetic spin wave mode. The AFM develops on cooling a distinctive shoulder (Fig 7(a)). FM mode has a broad asymmetric profile between 4 K and 25 K (Fig. 7 (b)).

We already commented that in $SmCrO_3$ a conventional Lorentzian fit of the ferro- and anti-ferromagnetic spin modes shows $T_{SR}$ dependency. Most important, understanding the two modes carries implicit moment dynamics. It is worth noting that the calculation resulting in eqs. (1, 3, 4) utilizes in the mathematical analysis the two-sublattice magnetic model for antiferromagnetics favored by isotropic superexchange interactions. This applies successfully to the properties of a magnetic material at low static fields and assumes negligible lattice distortions.

On the other hand, it is also known the Rare Earth ion size in distorted-perovskite compounds drives changes in the A site polyhedra. This varies appreciably across the lanthanide series. In orthoferrites, from Gd to Lu, (and against the lighter more 4*f* delocalized La-Sm) oxygen polyhedra, are so distorted that some of the potential nearest neighbor's in the oxygen cage are at second neighbors atomic distances.[62] This distortion triggers in simple nickelates charge disproportion producing intermixing sublattices and shifting the temperature at which a metal-insulator phase transition takes place.[63]

Within this framework, our replacing of Sm by Er implies a significant alteration of A site locally lowering symmetries and changing the Cr-O-Cr superexchange angle.[10] It is known that in $RCrO_3$ (R=Rare Earth) the average angle reduces from 180º as the ionic radius decreases along intrinsic structural distortions.[7] It enhances the effects the centrosymmetry breaking that is at the core of the multiferroicity.

We propose this disruption also prompts in $ErCrO_3$ an effective departure of the two-sublattice magnetic approximation. The expected two band momenta degeneracy no-longer holds, and consequently, the excitation for the antiferromagnetic mode, whose origin depends on anisotropies in the ***ac*** plane, will split into two very close features. These are identified as the band at ~42 cm$^{-1}$ and its shoulder at ~45 cm$^{-1}$ shown in Fig.7 (a).

The FM mode at 26 cm$^{-1}$, rather asymmetric and heavily temperature dependent, is related to out of phase oscillations of z components in the $\Gamma_2$ phase.

The two THz features, at 43 cm$^{-1}$ and 26 cm$^{-1}$, in the nominal $\Gamma_4$, behave as a pair of locked bands sharing on cooling softening and weakening against complementing hardening. The simultaneous change in relative intensity of the FM and AFM resonances suggests off-***ac*** plane exchanges.

Our measurements also point to an environment in which a relative weak external field would yield unambiguous changes on the origin of the mechanism for canted ferromagnetism since, now, the magneto-crystalline anisotropy contains four non-equivalent magnetic ions with magnetization slightly different even at low static fields.[64] This will add to the complexity of coexisting spectra from the six non-equivalent rare earth ion sites introducing lattice local distortions.

To probe on this we choose to apply an external magnetic field at 5 K and 2 K because these are the temperatures for different magnetic excitations in the $\Gamma_4$ and $\Gamma_1$ representation respectively.

Fig. 8(a) shows that both, FM and AFM excitations at 5 K, in the $\Gamma_4$ phase are well defined. Applying an external magnetic field introduces a disruption on the dominant exchanges but does not induce band split as found for both modes in $SmCrO_3$ (Fig. 4). Rather, there is an initial weak linear field dependence, with either band softening or hardening. Peak positions are shown in Fig. 9(a) and 9(b). The horizontal arrows points to where the applied field overcomes the net internal field of the ferromagnetic resonance. It is about the value where the induced spin reversal is observed and there is a crystal field level split yielding to the $\Gamma_2$ new magnetic phase (Fig 6, inset).

At about 43 cm$^{-1}$ (vertical arrow in Fig. 8(a)) the profile of the AFM changes by shoulder merging into a unique broader band that net increases its half width at half maximum. For fields above 3 T, and higher, the AFM then hardens while moving toward merging and disappearing into a growing background that itself, field induced, grows continuously from the start. The ferromagnetic FM mode at ~26 cm$^{-1}$ and 5 K although remains well defined also undergoes broadening under the applied field implying some kind of delocalization until merging into the continuum. We understand this last global feature as likely originated in out of equilibrium magnetic and electric dipoles (domain walls?) inhomogeneous configurations generating the strong unstructured absorbance consequence of their entanglement in a many body scenario. We speculate that this unexpected background is what precludes the detection of the field dependent split at each resonant energy.

At 2 K, Fig, 8(b), the ~42 cm$^{-1}$ AFM resonance vanishes signaling the transition to the magnetic phase described by the $\Gamma_1$ irreducible representation (it does not allow canted ferromagnetism). This is characterized by the Rare Earth moments fully aligned antiferromagnetic and where the weak ferromagnetism associated with Cr disappears.[48] By removing the $K_c$ anisotropy and setting to zero the canting angle in equations (3, 4) the resonances now merge into only one that appears as a band stiffening by 4 cm$^{-1}$ the FM peak position. The energy of the resulting mode would depend only on the forces along the antiferromagnetic axis, and although oversimplified, this approach allows a guide for the resulting excitation. Local distortions will certainly have a non-negligible contribution disrupting the vector diagram of Fig. S2. Ruling out further low temperature structural changes we may consider that the subtle lattice distortions identified at point group Cs still remain and will compound local displacements inducing electric dipoles associated to ferrolectricity.[46]

The FM band temperature dependence may be followed in Fig. 7(b). Starting at ~12 K the rather weak broader and asymmetric profile in the $\Gamma_4$ representation reflects the FM dependence on anisotropies. In turn, it acquires an almost perfect Lorentzian shape when moment fluctuations are minimized reaching the $\Gamma_1$ phase at 2 K.

We also found that the $\Gamma_2$ phase might be though as an intermediate continuous second order phase transition going from $\Gamma_1$ to $\Gamma_4$ being characterized by a β critical exponent. We found that adjusting a power law to the experimental data using

$$\omega_{soft} = A_{FM} \cdot (T_{SR}-T)^{\beta} \qquad (6)$$

being A a constant and $T_{SR}$ an effective critical temperature yields a three parameter power fit to the temperature dependent peak positions with $A_{FM} = 7.15 \pm 0.05$, $T_{Sr} = 15.2 \pm 0.8$ K and β=0.520±0.005. This corresponds to the Landau critical exponent for spontaneous magnetization in an ordered phase. It is, however, worth stressing that this mode behavior has the reorientation temperature and not the antiferromagnetic transition temperature as the critical temperature.

Overall, all features in the measured spectra of $ErCrO_3$ at 5 K and 2 K, Figs. 8(a) and 8 (b), broaden and weaken upon increasing fields until merge into the growing continuum. Figs. 9 (a) and 9(b) (upper panels) contrast in absolute values the absorption profiles in the THz at 0 T and 10 T.

## CONCLUSIONS

Summarizing, we discussed absorbance of spin wave resonant modes and crystal field excitations of $RCrO_3$ (R=Pr, Sm, Er) ceramics in the THz regime at and below the spin reorientation temperatures. We trace the no detection of the ferromagnetic and the antiferromagnetic modes in $PrCrO_3$ to $Pr^{3+}$ remaining paramagnetic disrupting $Cr^{3+}$-$Pr^{3+}$ exchange correlations and anisotropies up to our lowest measuring temperatures. Seeking to establish the momenta role in these chromites at the magnetic reversal temperature $T_{SR}$ we extended our studies to $SmCrO_3$ and $ErCrO_3$. While k~0 spin wave resonances in $SmCrO_3$ reproduce the sharp low temperature hardening and joint behavior reported for SmFeO$_3$ (Ref. **29**) the ErCrO$_3$ modes have a different concerted temperature and intensity trends. We trace the origin of these differences to the disrupting role introduced by the Rare Earth ion size in the polyhedral oxygen cage of the perovskite. Lattice site *A* lifting momenta degeneracy suggest the need to go beyond the two magnetic sublattice approximation in dealing with the antiferromagnetic dynamics. We have also verified that, as predicted by eq. (1) and eqs. (3, 4), external fields in SmCrO$_3$ induce attenuation and strong band broadening in an overall linear split. In ErCrO$_3$ same magnitude fields expose a more complex scenario comprising magnetic excitation softening and band merging from a different dynamical regime involving spin reversal in potentially field deformed orbitals (lattice). It has a remarkably absence of the external field dependence suggested by eq. (1).

Our measurements bring up an often neglected role of the Rare Earth ion size and 4f electron valence in understanding the low temperature spin dynamics in these compounds. Measurements for PrCrO$_3$ as well as SmCrO$_3$ and the more unorthodox behavior found in ErCrO$_3$ suggests that the lattice may further deform under stronger magnetic fields beyond the temperature driven deformations found by far infrared reflectivity in the magnetically ordered phase (Fig. S2).[3] This points to exchange interactions function of the interatomic spacing[64] appearing as a topological distortion of the oxygen polyhedral occupied by the Rare Earth likewise discussed for EuTiO$_3$ (Ref. **65**). We also emphasize that resonant modes are only detected at temperatures below the onset at which the Rare-Earth starts ordering.

Absorption of low energy crystal field transitions in each compound show agreement with earlier calculations. A linear split is found in non-Kramers and Kramers Rare-Earth ions at low applied fields that turns quadratic for non-Kramers Pr$^{3+}$. The low temperature field induced spin flip into the $\Gamma_4$ (G$_x$, A$_y$, F$_z$) phase in ErCrO$_3$ triggers at ~2.5 T a remarkable secondary

Zeeman split in the upper most level of the already field split Kramers doublet.

Our research suggests an environment of magnetic and electric dipoles globally entangled in the unstable framework of the already volatile perovskite lattice. Beingoxides, they are bound to have octahedral rearrangements with random orbital misoriented contributions strongly coupled to electronic induced mechanisms for colossal magnetoresistance or polar ordering involving orbital/charge and/or spin fluctuations.[66]

## ACKNOWLEDGEMENTS

NEM thanks the Helmholtz-Zentrum Berlin (HZB), Germany, for beamtime allocation and financial help. He is also grateful to the CNRS-C.E.M.H.T.I. laboratory and staff in Orléans, France, for research and financial support in performing far-infrared measurements.

The authors are also indebted with D. De Sousa Meneses (C.E.M.H.T.I.) for sharing his expertise on infrared techniques.

JAA acknowledges the Paul Scherrer Institute (Swizerland) for the allowed neutron time, and the financial support of the Spanish "Ministerio de Economia y Competitividad" (MINECO) to the project MAT2013-41099-R.

# REFERENCES

1.  Aleonard R., Pauthenet R, Rebouillat J. P., and Veyret C., 1968 J. of Appl. Phys. **39**, 379.
2.  Sahu J. R., Serrao C. R., Ray N., Waghmare U. V., and Rao C. N. R. 200 7 J. Mater. Chem. **17**, 42.
3.  Supplementary Information for the compounds structural and absorbance analyses (XYZ )
4.  Geller S., 1957 Acta Crystallogr. **10**, 243.
5.  Ghosh A., Dey K., Chakraborty M., Majumdar S., and Giri S., 2014 Eur. Phys. Letter **107,** 47012.
6.  Alonso J. A. on $NdCrO_3$, private communication.
7.  Mahana S., Rakshit B., Basu R., Dhara S., Joseph B., Manju U., Mahanti S. D., and Topwal. D.  arXiv 1706.005447v1
8.  Yamaguchi T., 1974 J. Phys. Chem Solids **35**, 479.
9.  Ray N., and Waghmare U. V., 2008 Phys. Rev. B **77**, 134112.
10. Zhou J.-S, Alonso J. A., Pomjakushin V., Goodenough J. B., Ren Y., Yan J.-Q., and Cheng J.-G., 2010 Phys. Rev. B **81**, 214115.
11. Talbayev, LaForge A. D., Trugman S. A., Hur N., Taylor A.J., Averitt R. D., and Basov D. N., 2008 Phys. Rev. Lett. **101**, 2476015.
12. Goodenough J. B., 2004 Rep. Prog. Phys. **67**, 1915.
13. Hornreich R. M., Komet Y., Nolan R., Wanklyn B. M., and Yaeger I., 1975 Phys. Rev. B **12**, 5094.
14. Tsushima K., Aoyagi K., and Sugamo S., 1970 J. Appl. Phys. **41**, 1238.
15. Qian X., Chen L., Cao S., Zhang J., 2014 Sol. St. Commun. **195,** 21.
16. Cooke H., Martin D.M., and Wells M. R., 1974 J. Phys. **C 7**, 3133.
17. Hasson A., Hornreich R. M., Komet Y., Wanklyn B.M., and Yaeger I., 1975 Phys. Rev. B **12**, 5051.

18. Doroshev V. D., Kharnachev A. S., Kovtun N. M., Soloviev E. E., Chervonenkis A. Ya., Shemyakov A. A., 1972 Phys. Status Solidi **B 51** K31.

19. Keffer F., and Kittel C., 1952 Phys. Rev. **85**, 329.

20. Moriya T., Magnetism Vol1, Eds. Rado G. T. and Suhl H. Academic Press, 1963.

21. Dzyalonshinkii I. E., Zh. 1957 Eksp.Teor. Fiz. 32, 1547.

22. Treves D., 1962 Phys. Rev. **125**, 1843.

23. Logemann R., Rudenko A. N., Katsnelson M. I., and Kirilyuk A 2017 J. Phys.: Condens. Matter **29** 335801.

24. White R.M., Nemanich R. J., and Herring C., 1982 Phys. Rev. B **25** 1822.

25. Herrmann G. F., 1964 Phys. Rev. **133**, A1334, Herrmann G. F. 1963 J. Phys. Chem. Solids **24**, 597.

26. Constable E., Therahertz spectroscopy of magnetic metal oxides, Ph.D. thesis, University of Wollongong, (2015), and references therein.(E. Constable E., Cortie D. L., Horvat J., Lewis R. A., Cheng Z., Deng G., Cao S., Yuan S., and Ma G., 2014 Phys. Rev. B **90**, 054413).

27. Koshizuka N., and Hayashi K.,1988 J. Phys. Soc. Jpn**. 57,**. 4418.

28. Koslov G.V., Lebedev S. P., Mukhr A. A., Prokhorov A. S., Fedorov L. V., 1993 IEEE Trans. on Magnetics, **29** 3443.

29. Fu X., Zeng X.,Wang D., Chi-Zhang H., Han J., and Cui T.J., 2015 Sci. Rep. **5**, 14777; 10.1038/srep14777.

30. Zhang K., Xu K, Liu X., Zhang Z., Jin Z., Lin X., Li B., Cao S., and Ma G., 2016 Sci. Rep. **6**, 23648; 10.1038/srep23648.

31. Holldack K., and Schnegg A., J., 2016 Large Scale Res. Facil **2**, A51

32. Bertaut E. F., and Forrat F., 1956 J. Phys. Rad. **17**, 129.

33. Gordon J. D., Hornreich R. M. , Shtrikman S. and Wanklyn B. M., 1976 Phys. Rev. **13**, 3012.

34. Pataud P., and Sivardière J., 1970 J. Phys. (Paris) **31**, 1017.

35. Bertaut E. F., Mareschal J., de Vries G., Algonard R., Pauthenet R., Rebouillat J., and Zarubickaa V., 1966 IEEE. Trans Mag **2**, 453.

36. Fabbris G. F. L., Tuning Electronic Correlation with Pressure (2014). Arts & Sciences Electronic Theses and Dissertations. Paper 371, Washington University Open Scholarship.

37. Podlesnyak A., Rosenkrnz S., Fauth F., Marti W., Scheci H. J., and Furrer A., 1994 J. of Phys. Cond. Matter 6 4099.

38. Bertaut E. F., Bassi G., Buisson G., Burlet P., Chappert J., Delapalme A., Mareschal J., Roult G., Aleonard R., Pauthenet R, and Rebouillat J. P., 1966 J. Appl. Phys. **37**, 1038.

39. Shamir N., Melamud M., Shaked H., and Shtrikman S., 1977 Physica B+C **90,** 217.

40. Cooke A. H., Martin D. M., and WellsM. R. 1974 J; Phys. C: Solid State Phys. **7**, 3133.

41. Macfarlane R. M. and Shelby R. M., in Modern Problems in Condensed Matter Sciences; edited by Kaplyanskii A. A. and Macfarlane R. M., Spectroscopy of Solids Containing Rare Earth Ions, Vol. 21 (Elsevier, Amsterdam, 1987) pp. 51-184).

42. Veissier L., Thiel C. W., Lutz T., Barclay P. E., Tittel W., and Cone R. L., 2016 Phys. Rev. B **94**, 205133.

43. Cao J., Vergara L. I., Musfeldt J. L., Litvinchuk A. P., Wang Y. J., Park S., and Cheong S.-W, 2008 Phys. Rev. Lett. **100**, 177205.

44. Ghosh A., Dey K., Chakraborty M., Majumdar S., and Giri S., 2014 Eur. Phys. Letter **107,** 47012.

45. El Amrani M., Zaghrioui M., Ta Phuoc V., Gervais F., and Massa, N.E., 2014 J . of Mag. and Mag. Materials **361,** 1.

46. Rajeswaram B., Khomskii D. J., Zvezdin A. K., Rao C. N. R., and Sundaresan A., 2012 Phys. Rev. B **86**, 214409.

47. Gorodetsky G., Hornreich R. M., Shaft S., Sharon B., Shaulov A., and Wanklyn, B. M.. 1977 Phys. Rev. B **16**, 515.

48. Bertaut E. F., in Magnetism, edited by Rado G. T. and Suhl H. Academic Press, New York, Vol. 3, p. 149 (1963).

49. Tsushima K., Aoyagi K., and Sugano S., 1970 J. of Appl. Physics **41**, 1238.

50. Knížek K., Novák P., Jirák Z., and de la Cruz C. R., 2014 Solid State Sciences **28**, 26.

51. Jirák Z., Hejtmánek J., Knížek K., Novák P., Šantavá E., and Fujishiro H.. 2014 J. of Appl. Phys. 115, 17E118.

52. Ma, Jie, "Neutron Scattering Study of Charge-Ordering in $R_{1/3}Sr_{2/3}FeO_3$ (R=La, Pr, Nd, Sm, and Y)" (2010). Dissertations Paper 11490, Iowa State University

53. Aring K. B., and Sievers A. J., 1970 J. Appl. Phys. **41**, 1197.

54. White R. L., 1969 J. Appl. Phys. **40**, 1061.

55. Holmes L., Eibschütz M., and Van Uitert L. G.. 1970 J. Appl. Phys. **41**, 1184.

56. Bertaut F. and Mareschal J., 1967 Sol. St. Communications **5,** 93.

57. Eibschutz M., Holmes L., Maita J. P., and Van Uitert L. G.. 1970 Sol. St. Communications **8**, 1815.

58. Meltzer R. S. and Moos H. W., 1970 J. of Appl. Physics **41**, 1240.

59. Toyokawa K., Kurita S., Tsushima K. K.,1979 Phys. Rev. B **19**, 274.

60. Kaneko M., Kurita S., and Tsushima K., 1977 J. Phys. C: Solid State Phys. **10,** 1979.

61. Courts R. and Hüfner H., 1975 Z. Phys. **B22**, 245

62. Marezio M., Remeika J. P., and Dernier P- D., 1970 Acta Cryst. **B26**, 2008.

63. Medarde M., Dallera C, Grioni M., Delley B., Vernay F., Mesot J., Sikora M., Alonso J. A., and Martínez-Lope M. J., 2009 Phys. Rev. B **80**, 245105.

64. Tiwari B., Surendra M.K., and Rao M. R. S., 2014 Mater. Res. Express **1,** 36102.

65. Guguchia Z., Keller H., Köhler J., and Bussmann-Holder A., 2012 J. Phys.: Condens. Matter **24,** 492201.

66. Keimer B. and Oles A. M., 2004 New J. Phys. **6**, 1367, and references therein.

# FIGURE CAPTIONS

**Figure 1.** Crystal field transition in $PrCrO_3$. (a) Low temperature dependent absorbance; (b) External magnetic field dependent absorbance at 3 K. The spectra have been vertically offset for better viewing. Inset: Absolute relative absorbance of $PrCrO_3$ at selected magnetic fields.

**Figure 2.** Main crystal field transition in $PrCrO_3$ for selected magnetic fields and Lorentzian deconvolutions at 3 K. Inset: Lorentzian fit peak positions of Zeeman split.

**Figure 3.** Temperature dependent absorbance of $SmCrO_3$ ferromagnetic and antiferromagnetic spin wave resonances. The spectra have been vertically offset for better viewing. Inset: Peak positions and power law fits (eqs. 5 and 6) for both modes.

**Figure 4.** $SmCrO_3$ antiferromagnetic and ferromagnetic spin wave resonances as function of the applied magnetic fields. (a) Field dependent absorbance (dashed lines are guides for the eye. Traces have been vertically offset for better viewing). (b) Peak positions for the absorbance shown in (a).

**Figure 5.** Crystal field transition in $ErCrO_3$ (a) Temperature dependent. (b) Magnetic Field dependent (dashed and dotted lines are guides for the eye. Traces have been vertically offset for better viewing).

**Figure 6.** Crystal field transitions absorbance of $ErCrO_3$ Lorentzian and Weibull absorbance deconvolutions for selected magnetic fields at 3 K. The interrogation mark at 10 T indicates the growing continuum upon applying

fields (see text). Inset: Peak positions of Zeeman splits. The arrow points at ~2.5 T to the spin reversal induced split at the upper most level into the → $\Gamma_2 \rightarrow \Gamma_4$ ($F_z$) phases. Start traces signal the possible field induced split at the lower energy level (full circle) of the Kramers doublet appearing as a broad band centered at ~55 cm$^{-1}$ (also, Fig. S5).

**Figure 7.** Ferromagnetic and antiferromagnetic spin modes of ErCrO$_3$ (a) Temperature dependent absorbance. Note that on cooling the relative intensity of the FM and AFM mode reverses. Inset: (a) AFM linear behavior; (b) detailed measurement of the ferromagnetic resonance intensity-frequency dependence. Dashed line is the power law fit, eq. (6), in the $\Gamma_2$ phase above 4 K.

**Figure 8.** Field dependent absorbance of spin wave resonances in ErCrO$_3$. **(a)** Field dependent AFM and FM modes at 5 K (the spectra have been offset for better viewing). Note that at about 8 T both resonant excitations merge into the edge of a broad continuum triggered by the applied field. The pointed line is intended as guide for the eye smoothing the interference sinusoidal pattern in the spectrum **(b)** Resonant mode (FM) within the $\Gamma_1$ irreducible representation at 2 K (the spectra have been offset for better viewing). Note that at about 8 T the resonant excitation (FM) merge with the edge of a broad continuum absorbance triggered by the applied field. The pointed line is intended as guide for the eye smoothing the interference sinusoidal pattern in the spectrum

**Figure 9.** **(a)** (upper panel) profile changes in the absolute absorbance at 0.0 T and 10 T at 5 K. The pointed line is intended as guide for the eye smoothing the interference sinusoidal pattern in the spectrum; (lower panel) Field dependent peak positions of the FM and AFM modes at 5 K. The arrows point to the field value at which the spin reversal takes place. **(b)** (upper panel) profile changes in the absolute absorbance at 0.0 T and 10 T at 2 K. The pointed line is intended as guide for the eye smoothing the interference sinusoidal pattern in the spectrum; (lower panel) Field dependent peak position of the only resonant mode left in the $\Gamma_1$ irreducible representation at 2 K. The arrow points to the field value at which the spin reversal takes place.

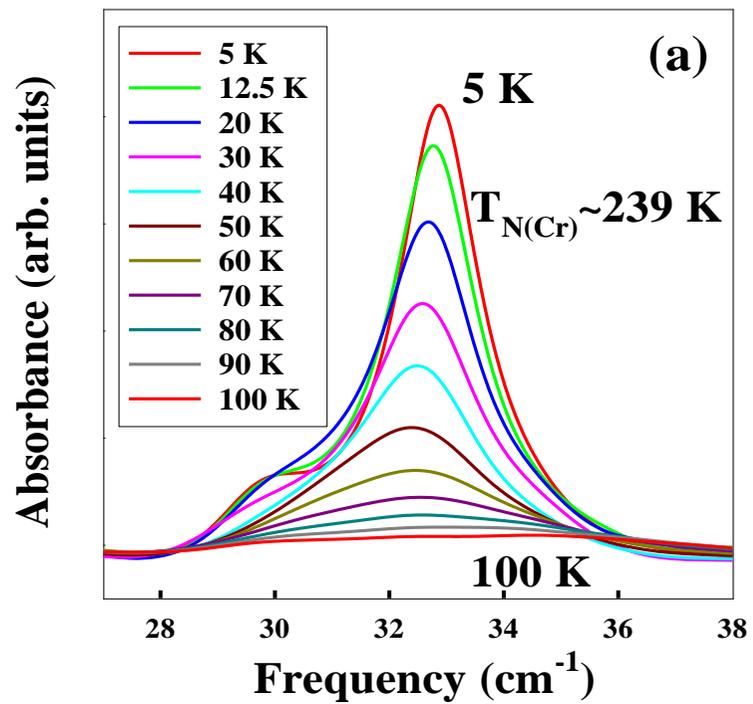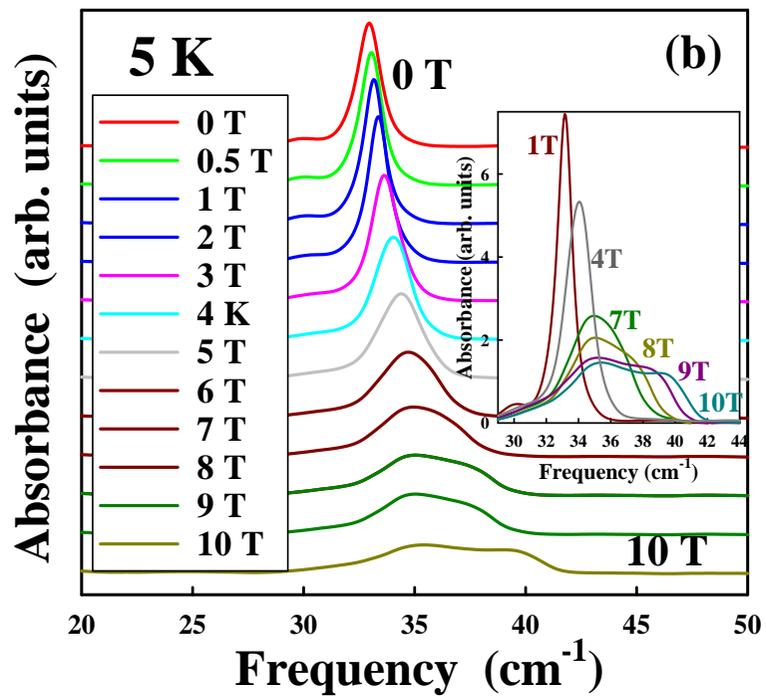

Figure 1
Massa et al

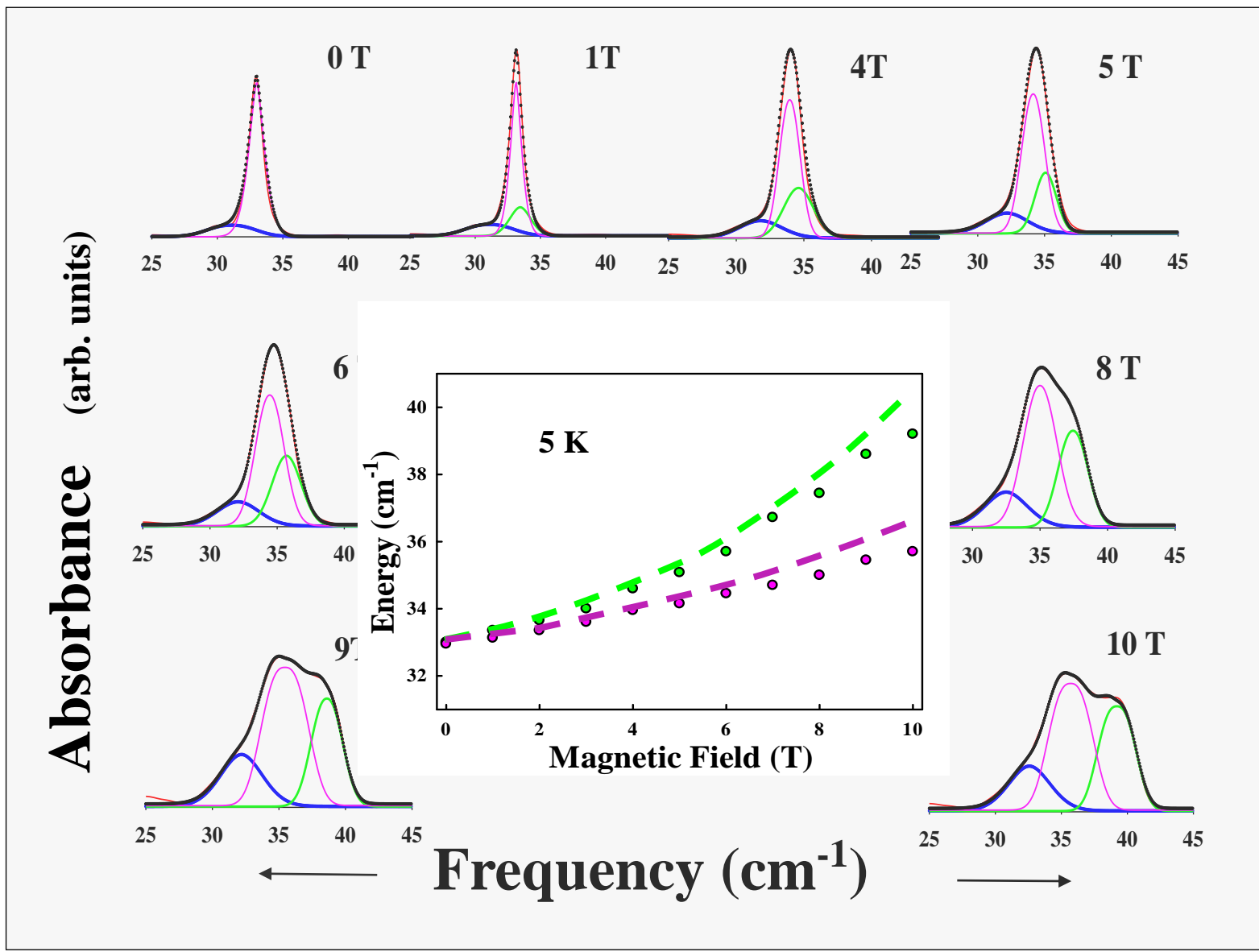

**Figure 2
Massa et al**

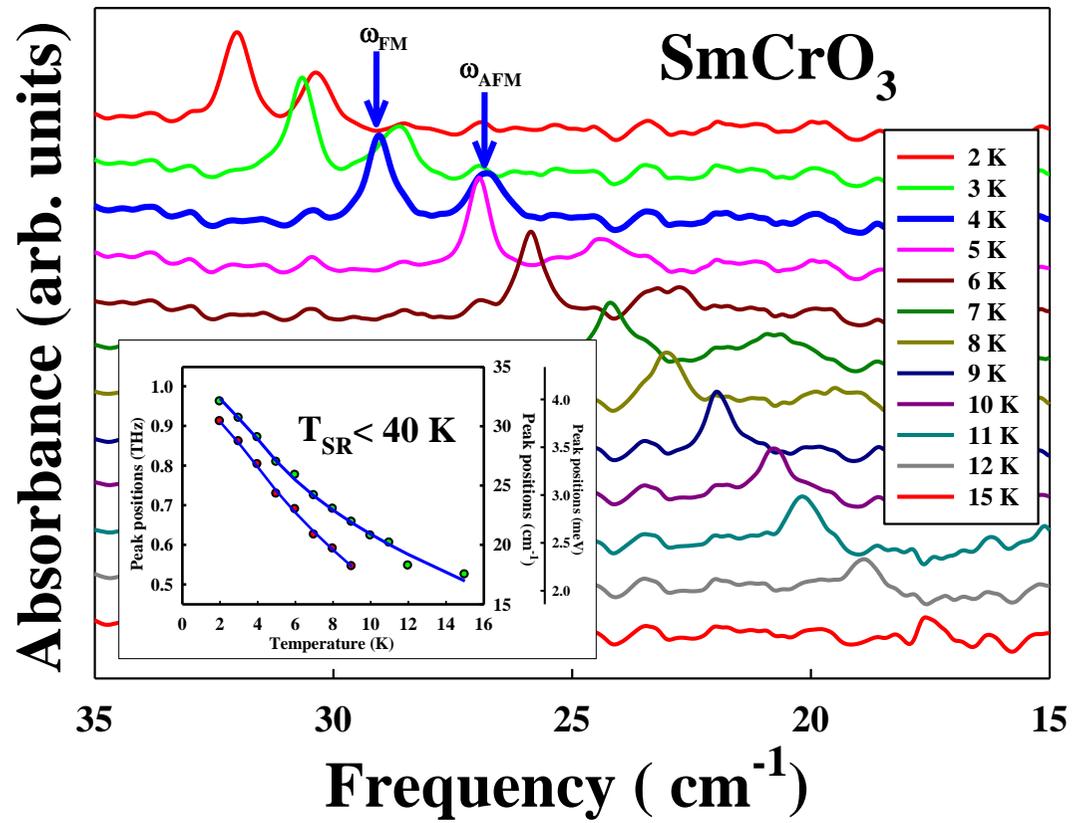

Figure 3
Massa et al

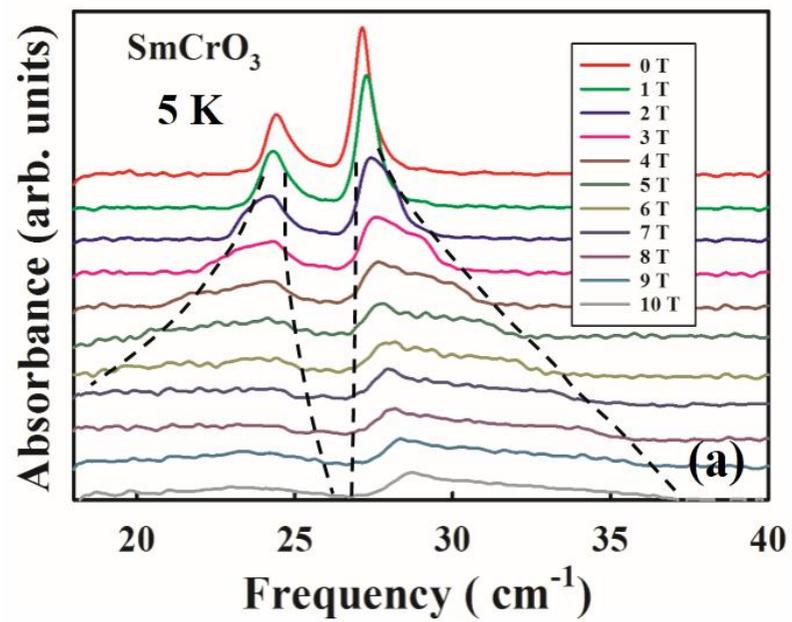 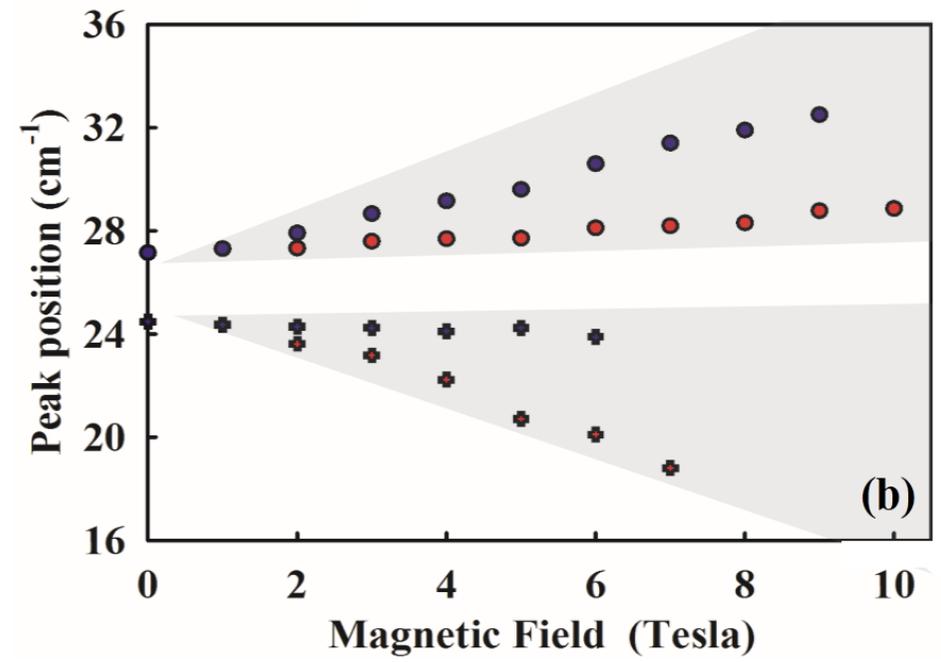

**Figure 4**

**Massa et al**

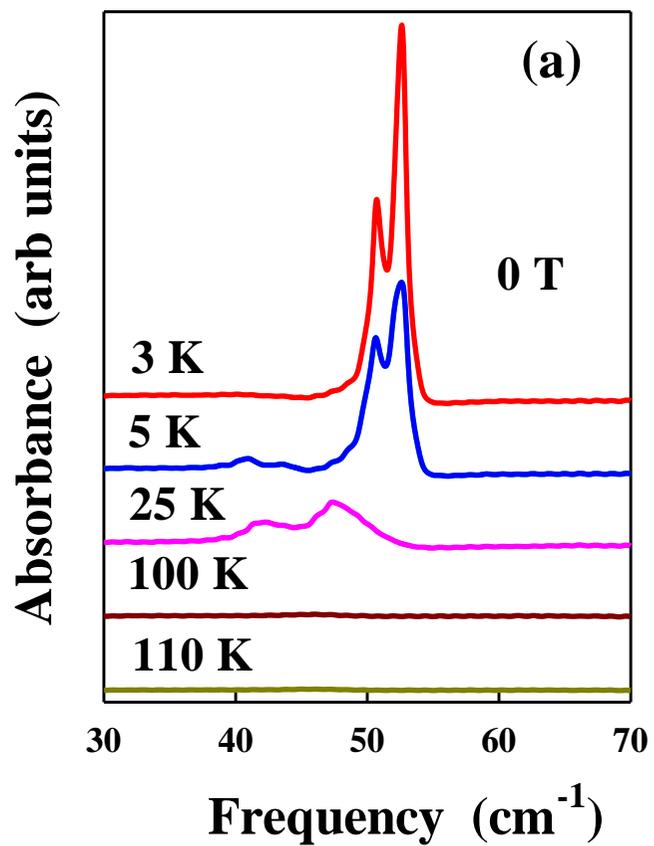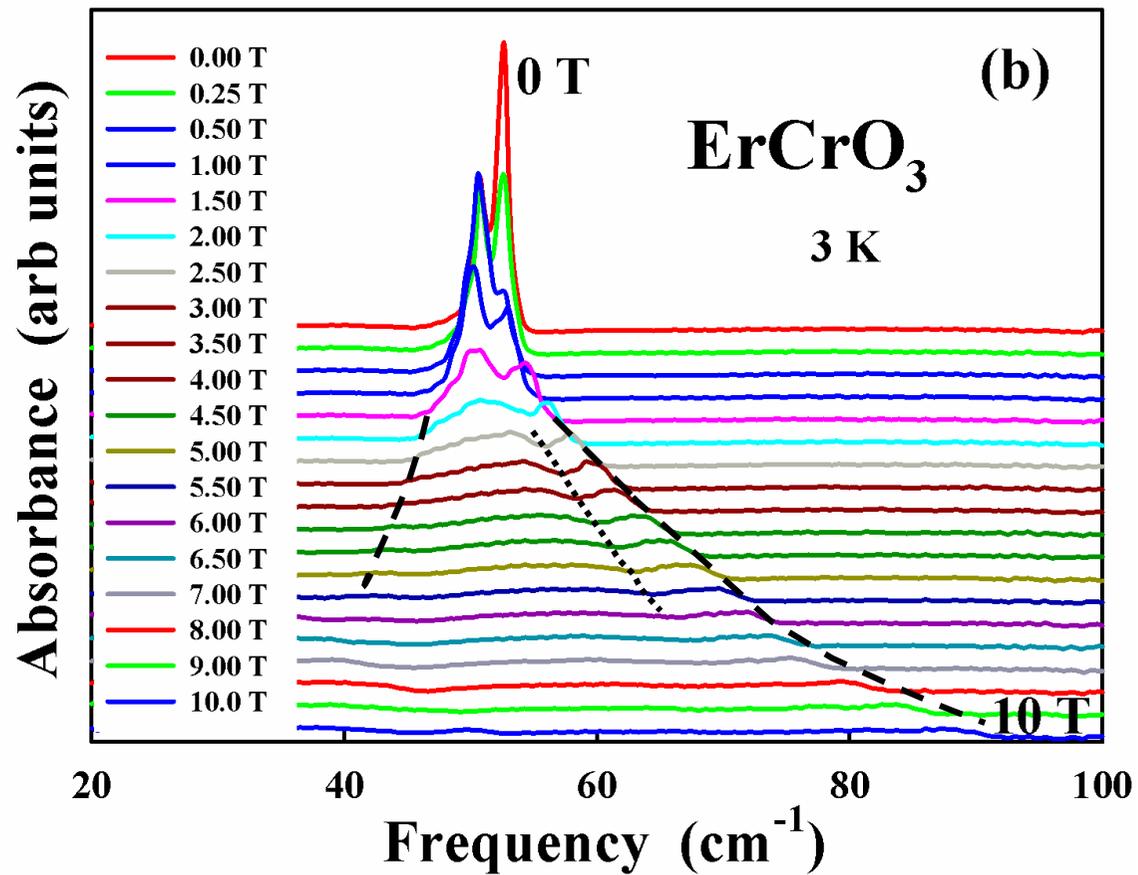

Figure 5
Massa et al

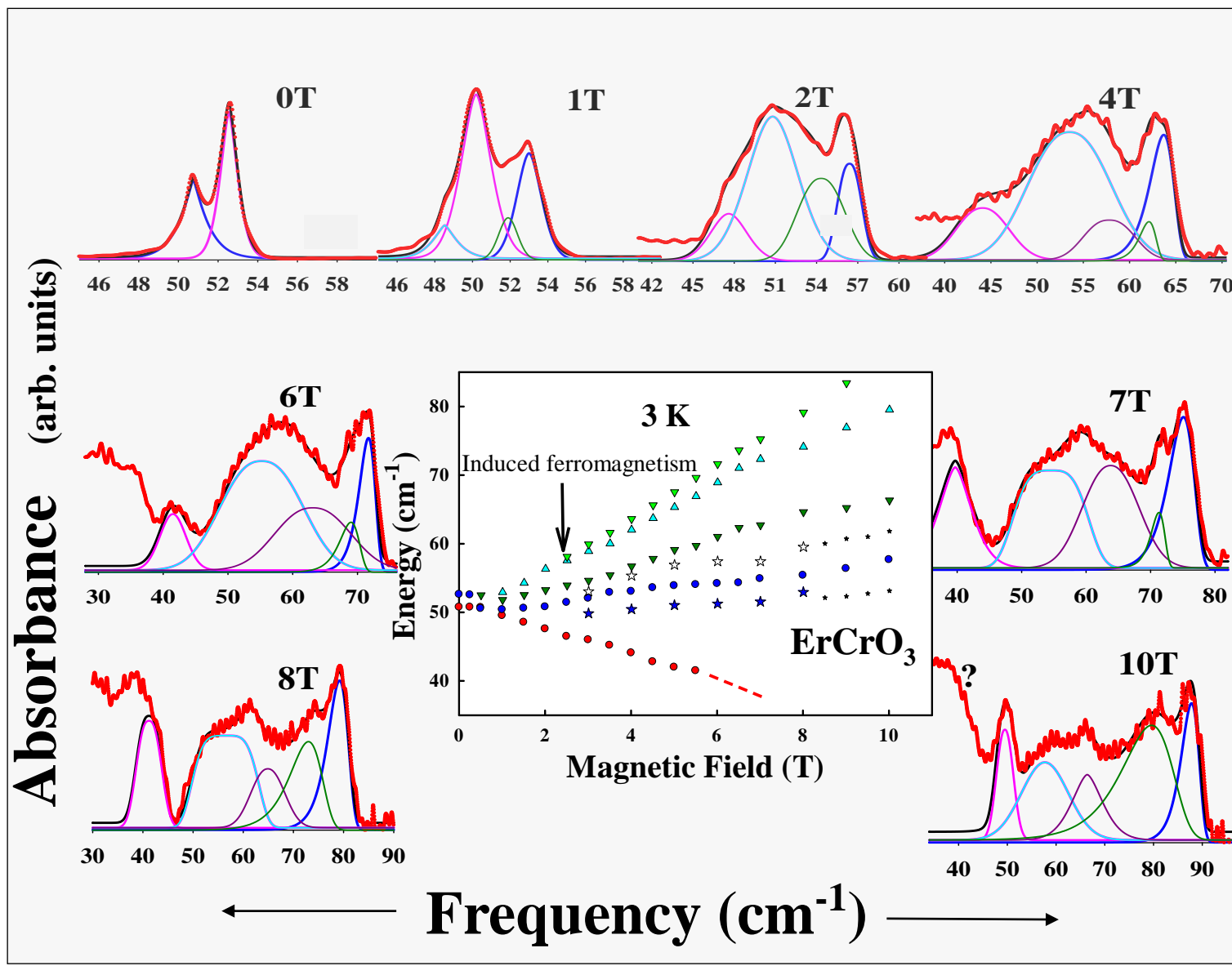

Figure 6
Massa et al

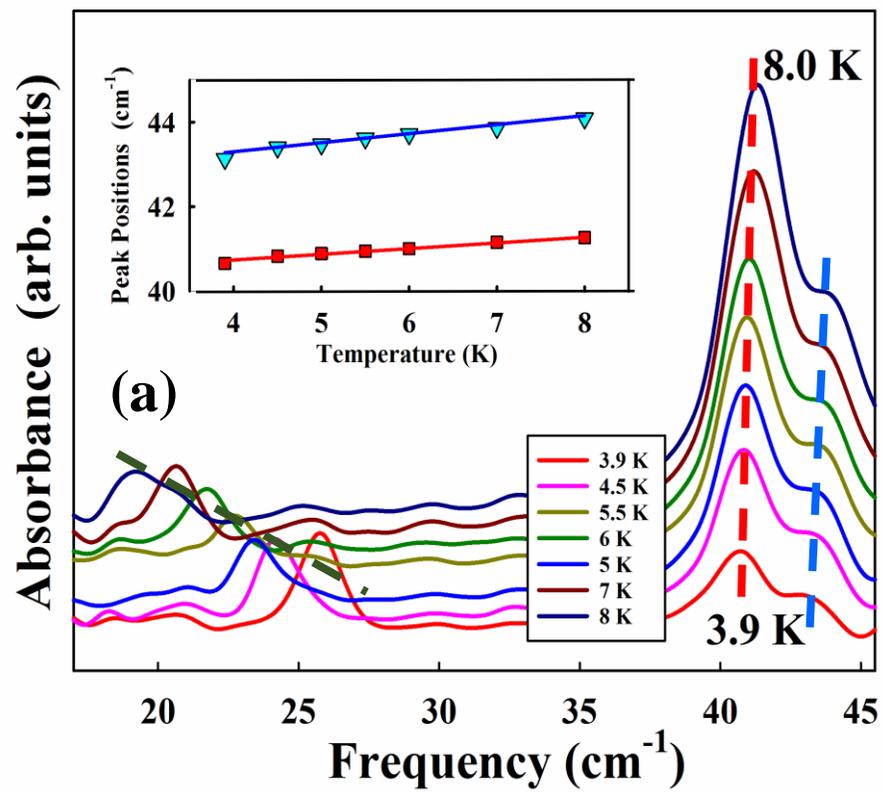 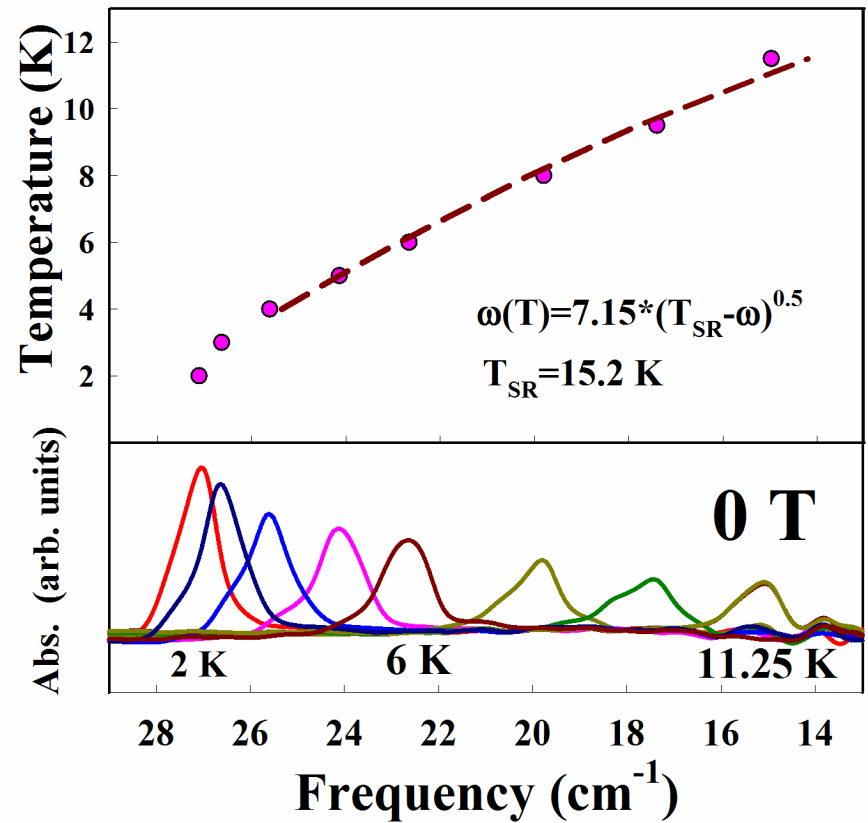

Figure 7
Massa et al

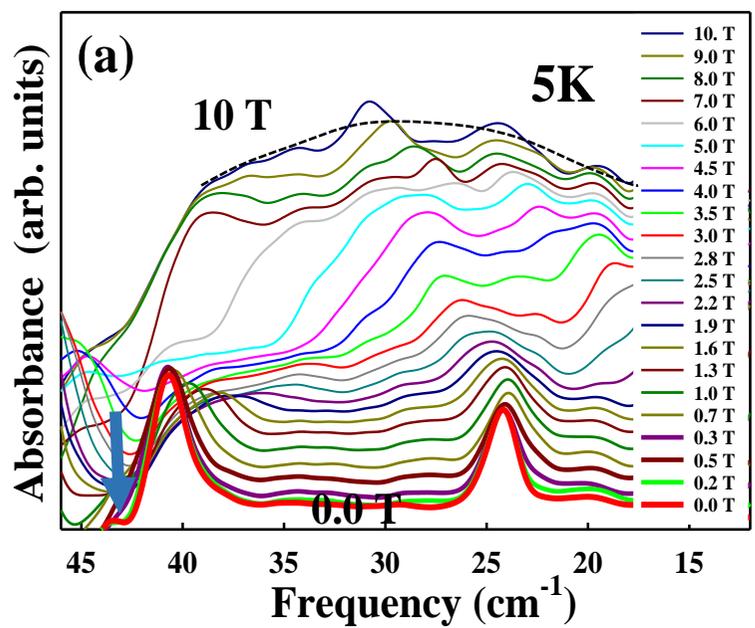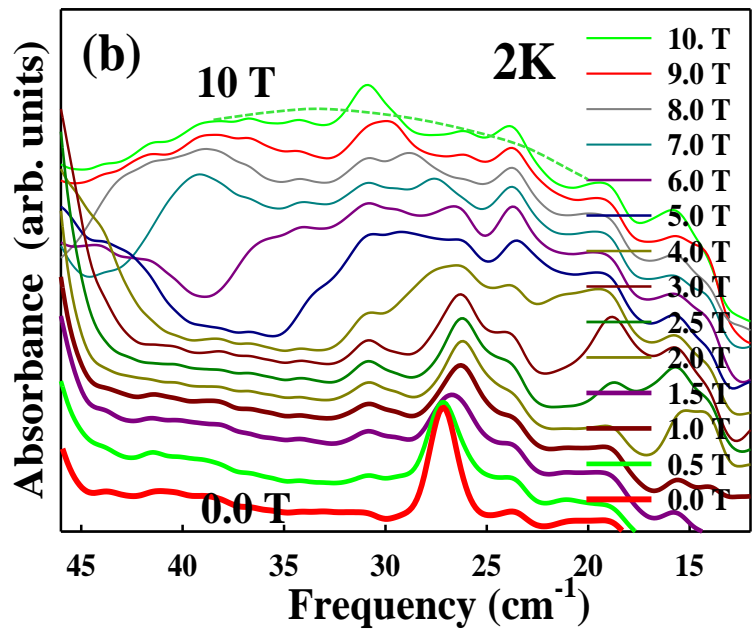

Figure 8
Massa et al

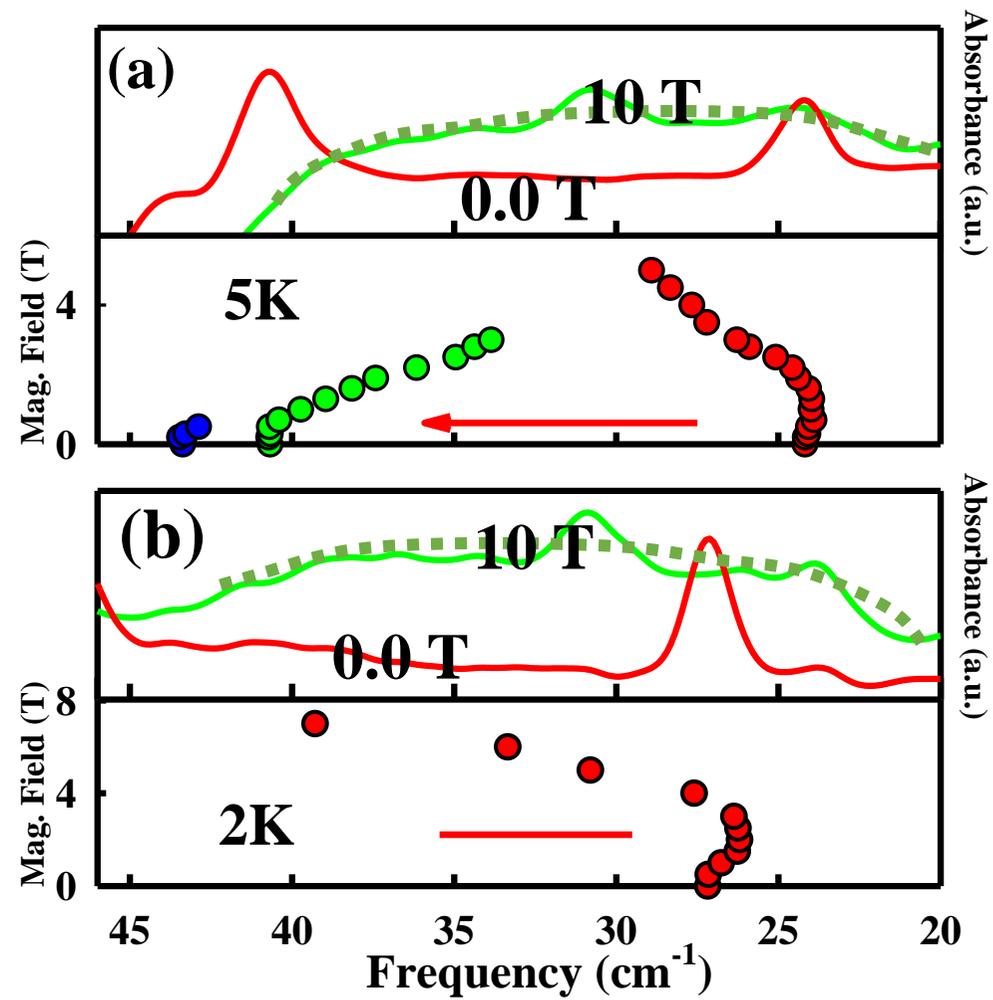

**Figure 9 Massa et al**

# Identification of spin wave resonances and crystal field transitions in simple chromites RCrO$_3$ (R=Pr, Sm, Er) at ultralow temperatures in the THz spectral region

Néstor E. Massa*,[1] Karsten Holldack,[2] Rodolphe Sopracase,[3] Vinh Ta Phuoc,[3] Leire del Campo,[4] Patrick Echegut,[4] and José Antonio Alonso[5]

[1] Laboratorio Nacional de Investigación y Servicios en Espectroscopía Óptica-Centro CEQUINOR, Universidad Nacional de La Plata, 1900 La Plata, Argentina.

[2] Helmholtz-Zentrum für Materialien und Energie GmbH, Institut für Methoden und Instrumentierung der Forschung mit Synchrotronstrahlung (BESSYII) Berlin, Germany.

[3] CNRS- Groupe de Recherche en Matériaux, Microélectronique, Acoustique et Nanotechnologies, Université François Rabelais. 37200 Tours, France.

[4] CNRS, CEMHTI UPR3079, Universté d' Orléans, F-45071 Orléans, France

[5] Instituto de Ciencia de Materiales de Madrid, CSIC, Cantoblanco, E-28049 Madrid, Spain.

•e-mail: neemmassa@gmail.com

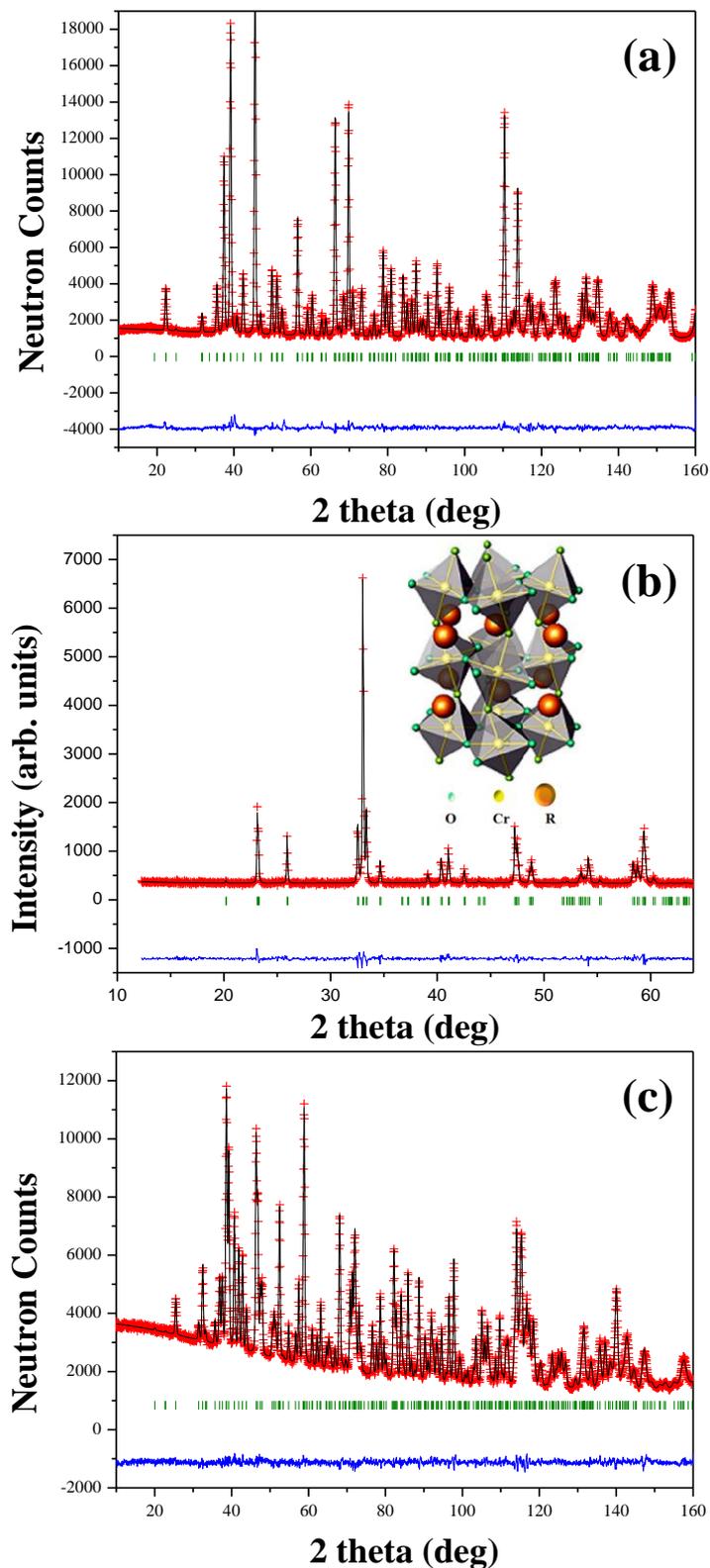

**Figure S1 (a)** Single phase neutron powder diffraction pattern of PrCrO$_3$ fitted at room temperature in the orthorhombic D$_{2h}^{16}$–P*bnm* space group; **(b)** single phase room temperature X-ray (CuKα) diffraction pattern for SmCrO$_3$ fitted at room temperature in the orthorhombic D$_{2h}^{16}$–P*bnm* space group. The inset corresponds to a view of the orthorhombic perovskite structure (O: oxygen; Cr: Chromium; R: Rare Earth); **(c)** Single phase neutron powder diffraction pattern of ErCrO$_3$ fitted at room temperature in the orthorhombic D$_{2h}^{16}$–P*bnm* space group.

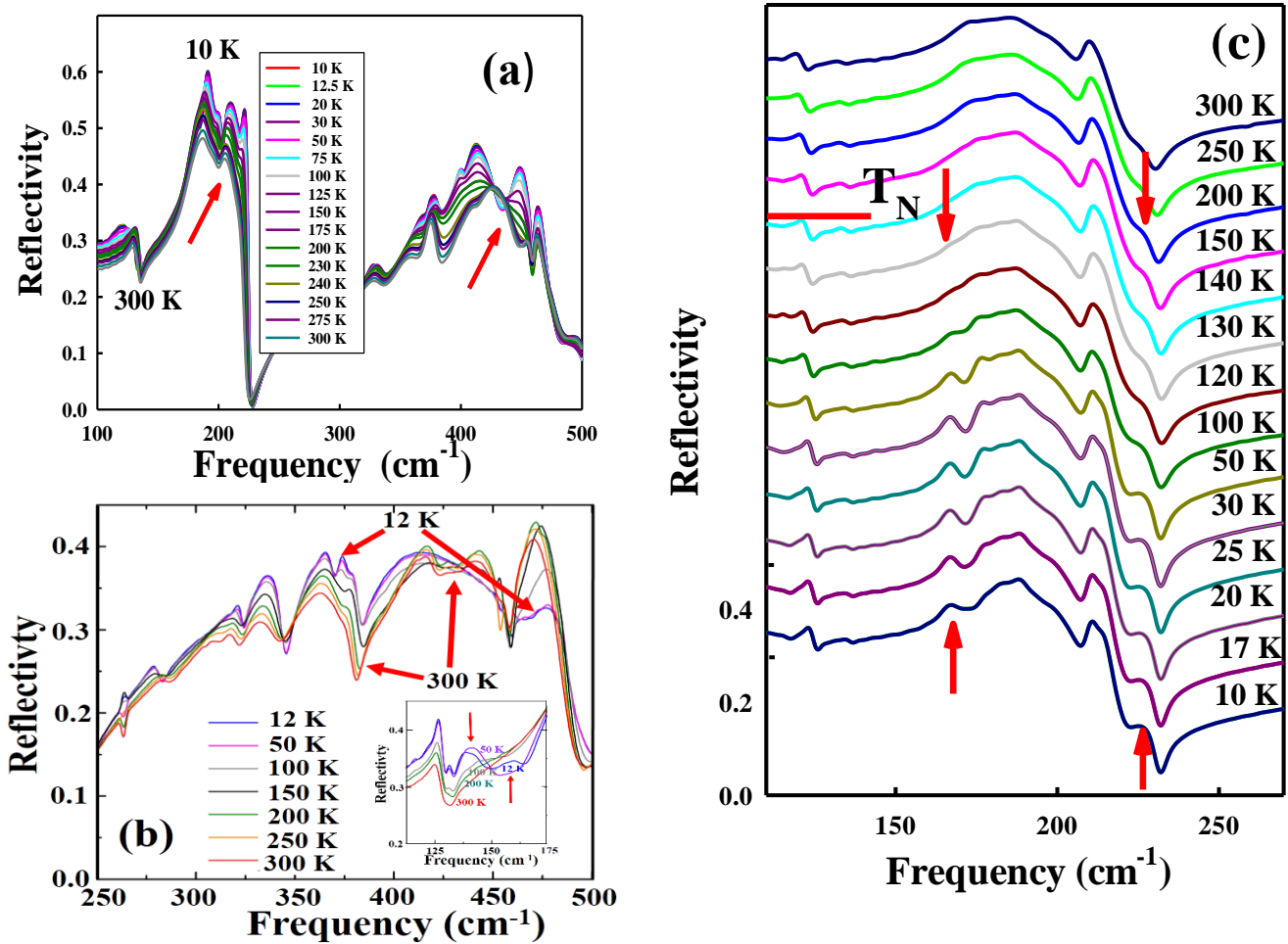

**Figure S2.** Far-infrared temperature dependent reflectivity from ambient to 10 K. Arrows point to emerging phonons as the sample cools down denoting subtle changes in symmetry. (a) $PrCrO_3$; (b) $SmCrO_3$ (after ref. 3); (c) $ErCrO_3$.

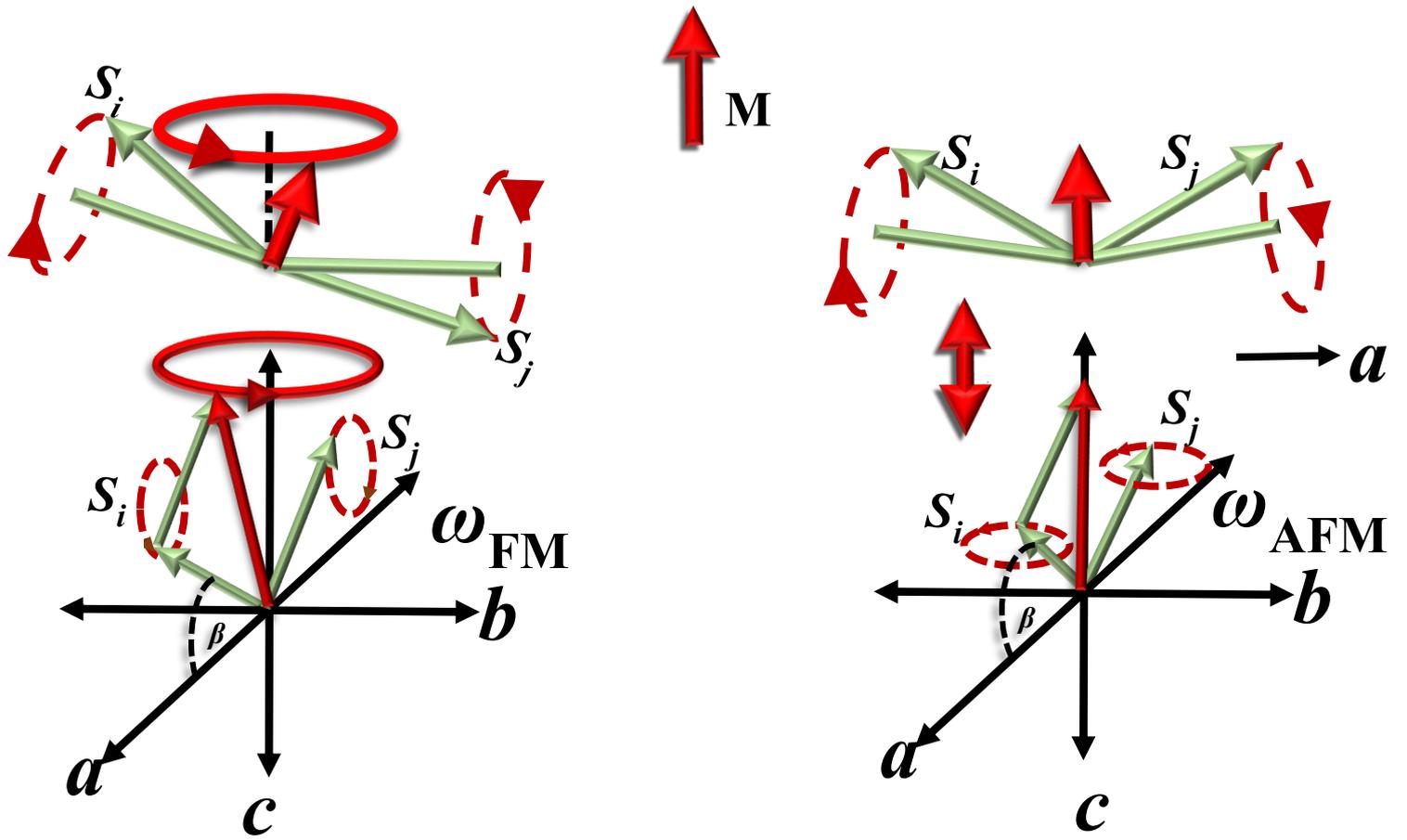

**Figure S3**. THz magnetic resonances of optically active spin wave modes in the $\Gamma_4$ representation of orthorhombic distorted perovskites. $S_i$ and $S_j$ are the transition metal spins along the ***a*** axis. The AFM mode is the amplitude oscillation and the FM mode is the precession of the ferromagnetic moment M. (After Lu et al (ref. 1) and Constable et al (ref. 2)).

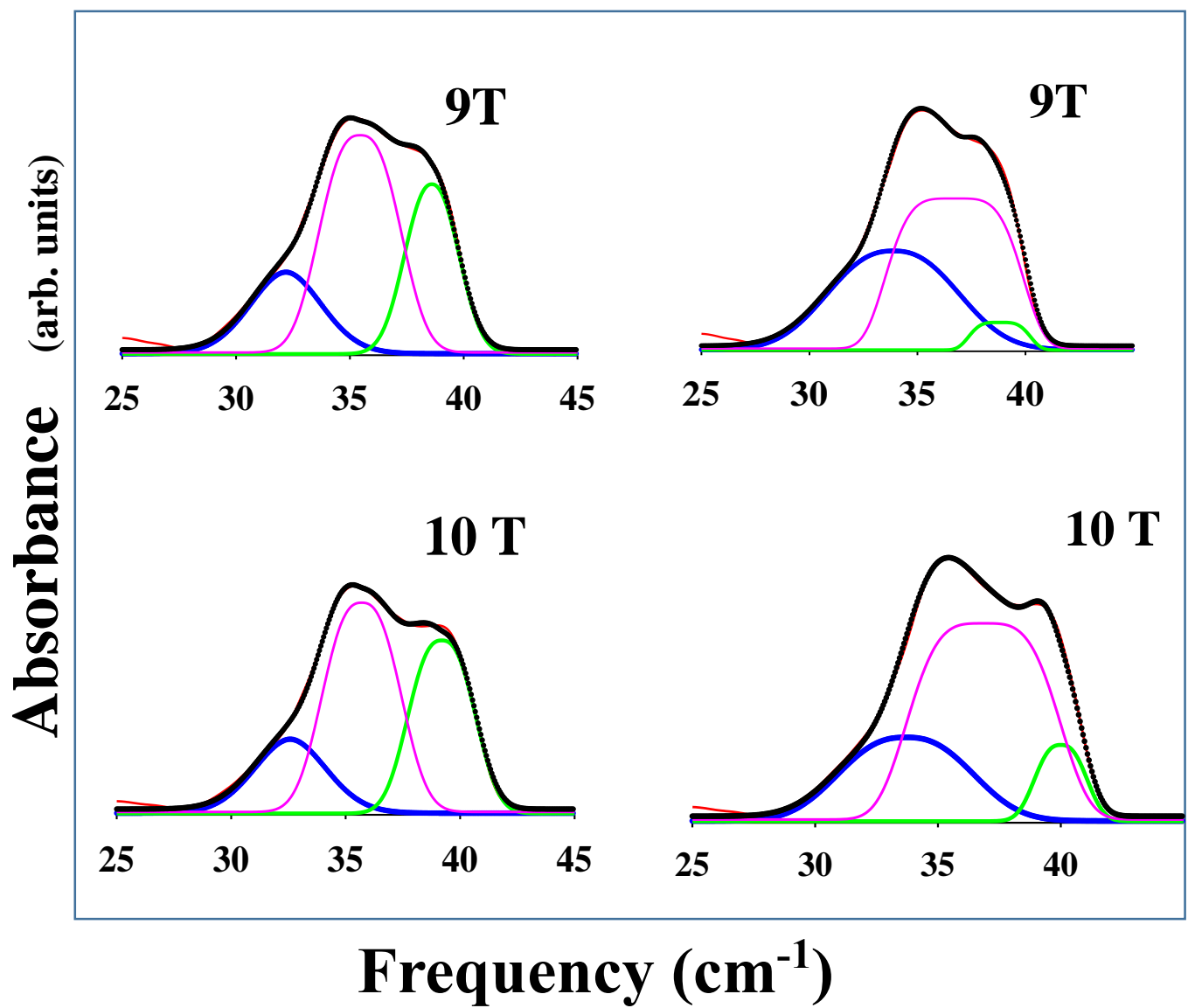

**Figure S4**. Deconvoluted absorbance spectra of PrCrO$_3$ at 5 K under the highest applied magnetic fields defining the lower continuous and upper dashed curves shown in Figure 2 (central inset) defining the confidence band.

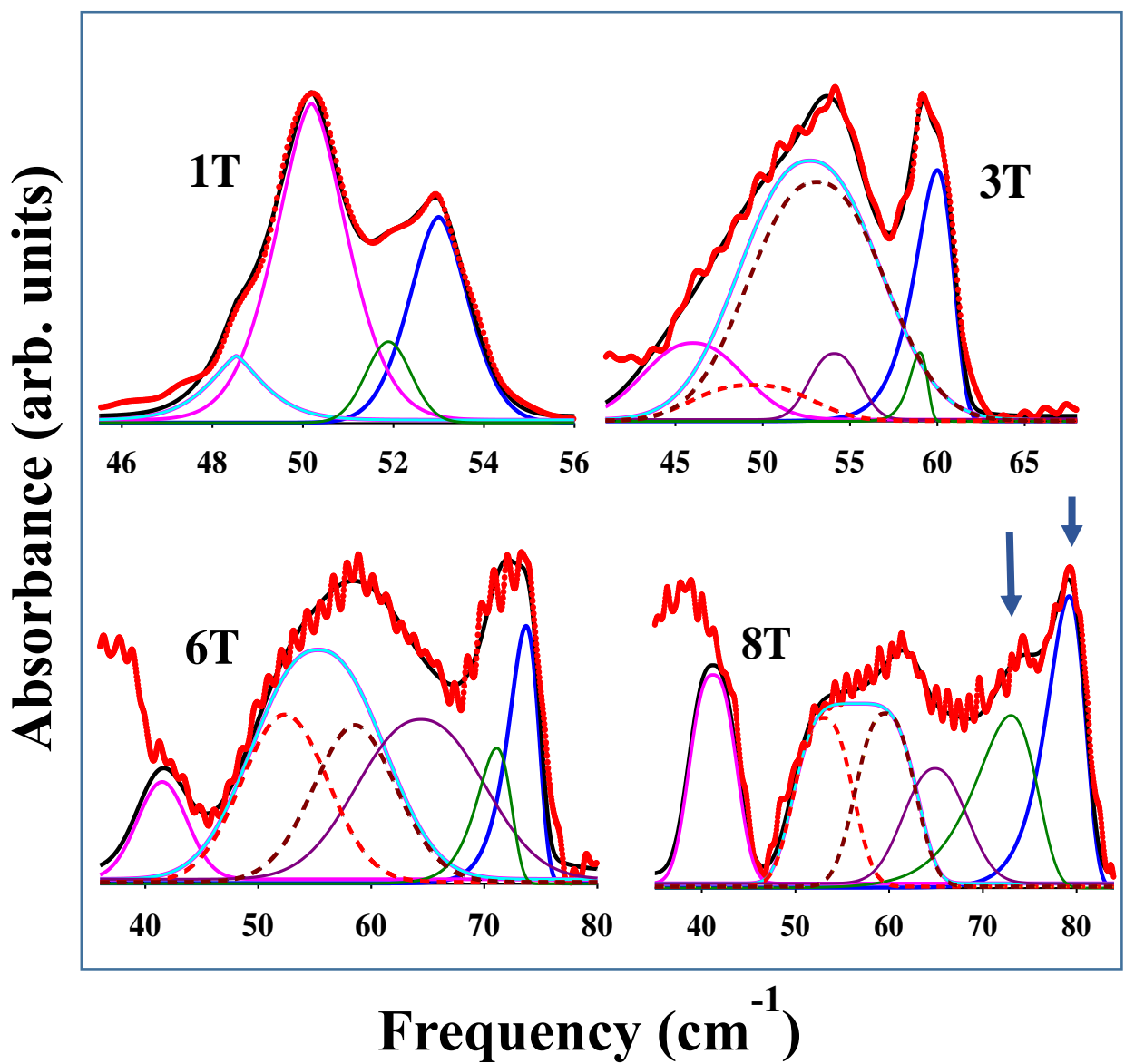

**Figure S5.** Detail on experimental main crystal field transition absorbance of ErCrO$_3$ at 3 K for selected magnetic fields and Lorentzian and Weibull deconvolutions. Arrows in the 8 T band profile point the secondary Zeeman split level triggered by the spin field induced reorientation → $\Gamma_2$ → $\Gamma_4$ (F$_z$). Dashed lines indicate field induced split at the lower energy level (full line envelope) of the Kramers doublet centered at ~55 cm$^{-1}$

# REFERENCES

1. Lu J, Li X., Hwang H. Y., Ofori-Okai B. K., Kurihara T., Suemoto T., and Nelson K. A., arXiv.org, 1605.06476.
2. E. Constable, D. L. Cortie, J. Horvat, R. A. Lewis, Z. Cheng, G. Deng, S. Cao, S. Yuan, and G. Ma, Phys. Rev. B **90**, 054413 (2014).
3. M. El Amrani, M. Zaghrioui,∗, V. Ta Phuoc, F. Gervais, and N. E. Massa, J. of Mag. and Mag. Materials **361,** 1 (2014).